\RequirePackage{fix-cm}
\documentclass{svjour3}                     % onecolumn (standard format)
\smartqed  % flush right qed marks, e.g. at end of proof
\usepackage{graphicx}
\usepackage[authoryear,square,sort&compress]{natbib}
\usepackage{amsmath}

\usepackage{bm}

 \newcommand{\ie}{\mbox{\it{i.e.}}}
 \newcommand{\eg}{\mbox{\it{e.g.}}}
 \newcommand{\va}{{\bm a}}
 
 \newcommand{\vv}{{\bm v}}
 \newcommand{\vx}{{\bm x}}
 
 \newcommand{\cf}{\mbox{\it{cf.}}}
 \newcommand{\cO}{{\cal O}}

 \newcommand{\vr}{{\bm r}}
 \newcommand{\wbar}{{\bar w}{}}
 \newcommand{\e}{\varepsilon}
 
 \newcommand{\bea}{\begin{eqnarray}}
  \newcommand{\ba}{\begin{eqnarray}}
 \newcommand{\eea}{\end{eqnarray}}
  \newcommand{\ea}{\end{eqnarray}}

  \newcommand{\be}{\begin{equation}}
 \newcommand{\ee}{\end{equation}}
\newcommand{\lt}{\left}
\newcommand{\ri}{\right}
\newcommand{\la}{\label}

\journalname{Space Science Reviews}
\newcommand{\grdist}{\delta{g}}
\newcommand{\Xref}[1]{\overline{#1}}
\newcommand{\Xest}[1]{\widetilde{#1}}
\newcommand{\Xsyn}[1]{{#1}^{\ast}}

\begin{document}

\title{High Performance Clocks and Gravity Field Determination%
\thanks{The research of J. {M\"uller} has been supported by the Collaborative Research Center SFB 1128 "Relativistic Geodesy and Gravimetry with Quantum Sensors" at Leibniz Universit\"at Hannover. The work of S.M. Kopeikin has been supported by the grant No. 14-27-00068 of the Russian Science Foundation (RSF).
We gratefully acknowledge financial support from Labex FIRST-TF, ERC AdOC (grant n◦ 617553) and EMRP ITOC (EMRP
is jointly funded by the EMRP participating countries within EURAMET and the European Union).
} 
}


\author{J. {M\"uller}\and D. Dirkx\and S.M. Kopeikin\and G. Lion \and I. Panet  \and G. Petit  \and  P.N.A.M. Visser}


\institute{J. {M\"uller} \at
              Leibniz Universit\"at Hannover,
							Institut f\"ur Erdmessung,
              Schneiderberg 50, 30167 Hannover,
              Tel.: +49 511 762 3362,
              Fax: +49 511 762 400,
              \email{mueller@ife.uni-hannover.de} \\ 
           \and
					    D. Dirkx \at
              Faculty of Aerospace Engineering,
              Delft University of Technology,
              Kluyverweg 1, 2629 HS, Delft, The Netherlands,
              Tel.: +31 15 2788866,
              Fax: +31 15 2785322,
              \email{D.Dirkx@tudelft.nl}\;, \\
              Joint Institude for VLBI ERIC, 
              PO Box 2, 7990 AA Dwingeloo, The Netherlands\\
					\and
					    S.M. Kopeikin  \at
              Department of Physics and Astronomy,
              223 Physics Bldg.,
              University of Missouri,
              Columbia, MO 65211-7010,
              \email{kopeikins@missouri.edu}\;, \\
              Siberian State University of Geosystems and Technologies, Plakhotny Street 10, Novosibirsk 	   
              630108, Russia\\
           \and
            G. {Lion} \at
              LNE-SYRTE, Observatoire de Paris,
              PSL Research University,
              CNRS, Sorbonne Universit\'es,
              UPMC Univ. Paris~06, LNE,
              61 avenue de l'Observatoire, F-75014 Paris, France, \\
              \email{Guillaume.Lion@obspm.fr} \\
           \and
						I. Panet \at
              LASTIG LAREG, IGN, ENSG, Univ Paris Diderot, Sorbonne Paris Cité,
              35 rue H\'el\`ene Brion, 75013 Paris, France,
              Tel.: +33 1 57 27 53 34,
              \email{isabelle.panet@ensg.eu} \\
           \and
           G. Petit  \at
              Bureau International des Poids et Mesures,
              Pavillon de Breteuil,
              F-92312 S\`evres Cedex FRANCE,
              \email{gpetit@bipm.org} \\
           \and
			     P.N.A.M. Visser \at
              Faculty of Aerospace Engineering,
              Delft University of Technology,
              Kluyverweg 1, 2629 HS, Delft, The Netherlands,
              Tel.: +31 15 2782595,
              Fax: +31 15 2785322,
              \email{P.N.A.M.Visser@tudelft.nl}	
}

\date{Received:          date / Accepted:         date}  % The correct dates will be entered by the editor

\maketitle

\abstract{Time measured by an ideal clock crucially depends on the gravitational potential and velocity of the clock according to general relativity.  Technological advances in manufacturing high-precision atomic clocks have rapidly improved their accuracy and stability over the last decade that approached the level of $10^{-18}$.  This notable achievement along with the direct sensitivity of clocks to the strength of the gravitational field make them practically important for various geodetic applications that are addressed in the present paper.  

Based on a fully relativistic description of the background gravitational physics, we discuss the impact of those highly-precise clocks on the realization of reference frames and time scales used in geodesy. We discuss  the current definitions of basic geodetic concepts and come to the conclusion that the advances in clocks and other metrological technologies will soon require the re-definition of time scales or, at least, clarification to ensure their continuity  and consistent use in practice. 

The relative frequency shift between two clocks is directly related to the difference in the values of the gravity potential at the points of clock's localization. According to general relativity the relative accuracy of clocks in $10^{-18}$ is equivalent to measuring the gravitational red shift effect between two clocks with the height difference amounting to 1 cm.  This makes the clocks an indispensable tool in high-precision geodesy in addition to laser ranging and space geodetic techniques.

We show how clock measurements can provide geopotential numbers for the realization of gravity-field-related height systems and can resolve discrepancies in classically-determined height systems as well as between national height systems. Another application of clocks is  the direct use of observed potential differences for the improved recovery of regional gravity field solutions. Finally, clock measurements for space-borne gravimetry are analyzed along with closely-related deficiencies of this method like an extra-ordinary knowledge of the spacecraft velocity, etc. For all these applications besides the near-future prospects, we also discuss the challenges that are related to using those novel clock data in geodesy.


}

\section{Introduction}
\label{section:intro}

The development of ultra-precise clocks opens the possibility for 
enhancing several geodetic applications, including an improved definition of
timescales, datum connecting and the observation of Earth's gravity
field. This paper addresses the possible impact of high-precision clocks   
on these applications. 

The definition of timescales has wide-ranging consequences for several
applications. For example, there are numerous questions associated with 
realizing a global reference {\it closest to the mean sea level}, see \eg\ 
a discussion in \citep{sanchez2012}. Work is still under way in working 
groups of the International Association of Geodesy (IAG) to provide 
conventional definitions and procedures for the realization of a Vertical 
Datum Standardization. 
%There is evidence of inconsistency in the present set of definitions. 

A major objective of geodesy is the determination of physical
heights, \ie\ those heights related to the gravity field. The use of 
ultra-precise clocks to this aim has been addressed in \cite{mai2013}, 
indicating that clocks with an accuracy of $10^{-18}$ allow for example 
datum connections at a level of 1 cm if combined with precise positioning 
by Global Navigation Satellite Systems (GNSS). Combination of high-precision
terrestrial clock and GNSS data with \eg\ space-borne gravimetric data 
might enhance spatial resolution and accuracy of existing global gravity
field models and associated definition of vertical datums.

Finally, this paper will address the possible use of high-precision
space-borne clocks for global gravimetry. This could be realized with 
clocks that have an ultra-high stability at short integration times
\citep{mayrhofer2012}. An assessment will be made of the
achievable gravity field retrieval performance by an efficient
error propagation tool, where the performance for ultra-precise
clocks will be compared with other observation techniques including
the already proved techniques of low-low Satellite-to-Satellite
Traking (ll-SST) and Satellite Gravity Gradiometry (SGG).

This paper is organized as follows. After introducing the relevant background on relativistic reference systems (Section~\ref{sec:cel-system}) and their relation to the commonly used International Terrestrial Reference System (ITRS) in Section~\ref{sec:ter-system}, a recap is provided of the required
fundamental equations of time dilation (Section~\ref{sec:sb_grav_time}). 
The impact of high-precision clocks on the definition of timescales
will be addressed in Section~\ref{section:time}. The expected benefit of using
high-precision clocks for terrestrial gravimetry will be
discussed in Section~\ref{section:terrestrialy}, including their use
for datum connection and height comparison (Section~\ref{subsec:datum}) as well as 
for enhancing regional gravity field solutions
(Section~\ref{subsec:highres_ground}). This will be followed by an assessment
of using high-precision clocks for space-borne gravimetry 
(Section~\ref{section:space-borne}). Finally, conclusions
will be drawn in Section~\ref{section:conclusions}.

\section{Relativistic Celestial Reference Systems}
\label{sec:cel-system}

Current theory of relativistic reference systems in the solar system has been formulated by  \citet{Kopeikin_1988} and \citet{BK_1989_1,BK_1989_2} and further developed by \citet{DSX_1,DSX_2,DSX_3}. It was adopted by the IAU 2000 as a standard for processing high-precise astronomical and geodetic observations \citep{soffel2003}. This theory provides precise theoretical definitions of the barycentric celestial reference system (BCRS) and the geocentric celestial reference system (GCRS) as well as the relations between them \citep{Kopeikin_2011_book}. The important distinction of the BCRS and GCRS from their Newtonian counterparts comes from general relativity which predicts the effects of relativistic contraction of (coordinate) spatial distances and time dilation caused by the relative orbital motion of GCRS with respect to BCRS and the presence of the gravitational field of Sun, Moon, and other planets. Because of these effects the spatial coordinates of the two reference systems experience periodic relative variations  due to the  Lorentz and Einstein contractions of the relative order of $10^{-8}$. The basic coordinate time scales for BCRS and GCRS (called TCB and TCG respectively) are also different due to the time dilation and gravitational redshift \citep{soffel2003}.

In what follows the small letters refer to BCRS, whereas the capital letters do to GCRS. The Greek letters $\alpha, \beta,...$ denote spacetime indices taking values $0,1,2,3$ with the index $0$ belonging to time coordinate. Roman letters will denote the indices taking values $1,2,3$ corresponding to the spatial coordinates only. Bold letters denote spatial vectors, ${\bm x}=\{x^i\}=\{x^1,x^2,x^3\}$, ${\bm X}=\{X^i\}=\{X^1,X^2,X^3\}$, etc.
The symbol $\cO(c^{-n})$ means that all residual terms of order $c^{-n}$ are neglected.

The theory of astronomical reference systems that is outlined below, is formulated in the first post-Newtonian approximation (PNA) of Einstein's theory of gravity though it can be extended to high-order approximations if necessary.
The PNA is a weak-field, slow-motion approximation with three small parameters: $\epsilon = (GM/c^2 R)^{1/2}$ with $M$ and $R$ being a characteristic mass and size of a body; $\epsilon_\omega = \omega R/c$ with $\omega$ being a rotational angular velocity of the body, and $\epsilon_v=v/c$ where $v$ is the orbital velocity of the body. According to the virial theorem of gravitational physics \citep{goldstein_2002} these small parameters in the solar system are ordered as follows: $\epsilon_\omega < \epsilon_v<\epsilon$, and for simplicity one is using a common denominator of the small parameters, $c^{-1}$, as a book-keeping parameter for a post-Newtonian expansion of the metric tensor, although it is not dimensionless. The first post-Newtonian approximation neglects
terms of the order of $c^{-6}$ in the time-time component of the metric tensor $g_{00}$, terms of the order of $c^{-5}$ in the time-space components $g_{0i}$, and terms of the order of $c^{-4}$ in the space-space components $g_{ij}$. 

In Einstein's theory of gravity the gravitational field is described by a metric tensor denoted as $g_{\alpha\beta}=g_{\alpha\beta}(x^\mu)$ in BCRS and $G_{\alpha\beta}=G_{\alpha\beta}(X^\mu)$ in CGRS,  that also characterizes the geometry of spacetime. The metric tensor is found by solving the Einstein field equations which preserve their form irrespectively of the choice of coordinates. Thus, they can determine the components of the metric tensor only up to four degrees of freedom. This gauge freedom of the metric tensor components corresponds to the free choice of the coordinate system.
We will use the harmonic gauge condition \citep{Fock_1964}  imposed on the metric tensor in every coordinate system. The harmonic coordinates are convenient from a mathematical point of view and are often used in other branches of gravitational physics. 

Post-Newtonian form of the metric tensor in harmonic coordinates reads \citep{BK_1989_1,BK_1989_2, DSX_1}
\begin{align}\label{pnmetric}
g_{00} &= -1+\frac{2w}{c^2}-\frac{2w^2}{c^4}+\cO(c^{-6})\,& ,\quad \quad& G_{00} &=& -1+\frac{2V}{c^2}-\frac{2V^2}{c^4}+\cO(c^{-6}) \, , \nonumber \\
g_{0i} &= - \frac{4w^i}{ c^3}+\cO(c^{-5})  \, &,\quad\quad& G_{0i} &=& - \frac{4V^i }{c^3}\cO(c^{-5}) \, , \\
g_{ij} &=\phantom{-} \delta_{ij}\left(1+\frac{2w}{c^2}\right)+\cO(c^{-4})\,&,\quad \quad &G_{ij} &=&\phantom{-} \delta_{ij}\left(1+\frac{2V}{c^2}\right)+\cO(c^{-4}) \, , \nonumber
\end{align}
where the gravitational potentials are functions of time and spatial coordinates: $w\equiv w(t,{\bm x})$, $w^i\equiv w^i(t,{\bm x})$ in BCRS, and $V\equiv V(T,{\bm X})$, $V^i\equiv V^i(T,{\bm X})$ in GCRS, respectively. Notice that we have used a letter $V$ to denote the gravitational potentials of the geocentric metric tensor $G_{\alpha\beta}$. The paper by \citet{DSX_1} used $W$ instead of $V$. However, the letter $W$ is ubiquitously used in geodesy to denote the potential of the gravity force in rotating geocentric coordinates system and is represented as an algebraic sum of the gravitational potential $V$ and the centrifugal potential, $Z=1/2{\bm v}^2$ where ${\bm v}=\{v^i\}$ is velocity of observer with respect to GCRS (see, for example, Eq.~(\ref{kop5}) below.)
%%%%%%%%%%%%%%%%%%%%%%%%%%%%%%%%%%%%%%%%%%%%%%%%

The gravitational potentials are expressed in terms of the integrals taken over three-dimensional volumes occupied by matter of the bodies comprising the solar system. We single out the gravitational potentials associated with Earth and those produced by the external bodies (Moon, Sun, planets, etc.). In the post-Newtonian  approximation this decomposition reads
\bea
w(t,{\bm x})&=&w_E(t,{\bm x})+\bar w(t,{\bm x})\;,\\
w^i(t,{\bm {\bm x}})&=&w^i_E(t,{\bm x})+\bar w^i(t,{\bm x})\;,
\eea
and
\bea\label{dec321}
V(T,{\bm X})&=&V_E(T,{\bm X})+\bar V(T,{\bm X})\;,\\
V^i(T,{\bm X})&=&V^i_E(T,{\bm X})+\bar V^i(T,{\bm X})\;,
\eea
where the potentials with sub-index $E$ belong to Earth, and those with the bar to the external bodies. 

Explicit expressions of the geopotentials are given in the form of a particular solution of the Poisson equations that are volume integrals,
\bea
w_E(t,{\bm x})&=&G\int_{{\cal V}_E}\frac{\sigma(t,{\bm x}\rq{})d^3x\rq{}}{|{\bm x}-{\bm x}\rq{}|}+\cO(c^{-2})\;,\\
w^i_E(t,{\bm x})&=&G\int_{{\cal V}_E}\frac{\sigma^i(t,{\bm x}\rq{})d^3x\rq{}}{|{\bm x}-{\bm x}\rq{}|}+\cO(c^{-2})\;,\\
V_E(T,{\bm X})&=&G\int_{{\cal V}_E}\frac{\Sigma(T,{\bm X}\rq{})d^3X\rq{}}{|{\bm X}-{\bm X}\rq{}|}+\cO(c^{-2})\;,\\
V^i_E(T,{\bm X})&=&G\int_{{\cal V}_E}\frac{\Sigma^i(T,{\bm X}\rq{})d^3X\rq{}}{|{\bm X}-{\bm X}\rq{}|}+\cO(c^{-2})\;,
\eea 
where $\sigma(t,{\bm x})$, $\Sigma(T,{\bm X})$ and $\sigma^i(t,{\bm x})$, $\Sigma^i(T,{\bm X})$ are the post-Newtonian mass and mass-current densities of Earth\rq{}s matter in the BCRS and GCRS respectively. The densities are directly related to the model of the energy-momentum tensor of matter distribution inside Earth. In most of the practical applications considered below it is sufficient to assume that the mass densities are approximately equal
\be
\Sigma(T,{\bm X})=\sigma(t,{\bm x})=\rho(t,{\bm x})+{\cal O}\left(c^{-2}\right)\;,
\ee
where $\rho(t,{\bm x})$ is the baryon rest mass density. The current densities
\ba
\Sigma^i(T,{\bm X})&=&\Sigma(T,{\bm X})\left({\bm\omega}\times{\bm X}\right)^i\;,\\
\sigma^i(t,{\bm x})&=&\rho(t,{\bm x})\left[v^i_E+\left({\bm\omega}\times{\bm x}\right)^i\right]\;,
\ea
 where $v^i_E=v^i_E(t)$ and ${\bm\omega}=\{\omega^i(t)\}$ are the orbital velocity of Earth and the instantaneous angular velocity of Earth\rq{}s rotation, respectively, and the cross between two vectors denote the standard Euclidean cross product.
 
Earth\rq{}s gravitational potentials $V_E$ and $V_E^i$ admit multipolar expansion in the exterior space (outside the Earth),
\ba
V_E(T,{\bm X})&=&\frac{GM_E}{r}+\frac{GI_E^i X^i}{r^3}+\frac32\frac{GI_E^{ij}X^iX^j}{r^5}+{\cal O}\left(\frac{R^3_E}{r^4}\right)\;,\\
V^i_E(T,{\bm X})&=& \frac{G\dot I^i_E}{r}+\frac{G\left({\bm S}_E\times{\bm X}\right)^i}{2r^3}+\frac{\dot I^{ij}_EX^j}{2r^3}+{\cal O}\left(\frac{R^3_E}{r^3}\right)\;,
\ea
where $r=|{\bm X}|$ is the radial distance in GCRS, $M_E$ is the total (relativistic) mass of Earth, ${\bm S}_E=\{S^i_E\}$ is the angular momentum (spin) of Earth,  $I_E^i$ and $I^{ij}_E$ are dipole and quadrupole moments of Earth\rq{}s, and the overdot denotes a time derivative with respect to $T$=TCG. Usuall assumption is that the Earth\rq{}s center of mass is located at the origin of GCRS which makes $I^i_E=0$, and all terms depending on the dipole moment $I^i_E$ vanish from equations. We prefer to leave the dipole term explicitly in the equations because it is used as a vector parameter in tracking down the motion of the geocenter with respect to the origin of the International Terrestrial Reference System (ITRS) - for more detail see \citep{WuRay2012,Kuzin2010AdSpR,Altamimi2013AGUFM}.

Explicit expressions of the external gravitational potentials in BCRS are given in terms of the integrals performed over the volumes of the external bodies,
\bea
\bar w(t,{\bm x})&=&G\sum\limits_{A\not=E}\int_{{\cal V}_A}\frac{\sigma(t,{\bm x}\rq{})d^3x\rq{}}{|{\bm x}-{\bm x}\rq{}|}+\cO(c^{-2})\;,\\
\bar w^i(t,{\bm x})&=&G\sum\limits_{A\not=E}\int_{{\cal V}_A}\frac{\sigma^i(t,{\bm x}\rq{})d^3x\rq{}}{|{\bm x}-{\bm x}\rq{}|}+\cO(c^{-2})\;,
\eea
where $\sigma(t,{\bm x}\rq{})$ and $\sigma^i(t,{\bm x}\rq{})$ describe the distribution of mass and mass-current densities inside the volume of the external body $A$.

External gravitational potentials in GCRS are found as general solutions of the Laplace homogeneous equation that are given in terms of polynomials 
\bea\label{tidepot1}
\bar V(T,{\bm X})&=&Q_i X^i+\frac12 Q_{ij}X^i X^j+\frac16Q_{ijk}X^iX^jX^k+{\cal O}\left(X^4\right)\;,\\
\bar V^i(T,{\bm X})&=&C_{ij}X^j+\frac12C_{ijk}X^jX^k+{\cal O}\left(X^3\right)\;,
\eea 
where $Q_i=Q_i(T)$, $Q_{ij}=Q_{ij}(T)$ and $Q_{ijk}=Q_{ijk}(T)$ are the dipole, quadrupole and octupole moments of the tidal {\it gravitoelectric} field, $C_{ij}=C_{[ij]}(T)$, $C_{ijk}=C_{[ij]k}(T)$ are the quadrupole and octupole moments of the tidal {\it gravitomagnetic} field where the square parentheses around indices indicate the antisymmetry. The external multipole moments can be expressed in terms of the partial derivatives from the external gravitational potentials which are found by making use of the asymptotic matching of the metric tensor in BCRS and GCRS \citep{Kopeikin_2011_book}
\bea
Q_{ij}&=&\bar w_{,ij}({\bm x}_E)\;,\\
Q_{ijk}&=&\bar w_{,ijk}({\bm x}_E)\;,\\
C_{ijk}&=&\frac43\left[v_E^{[i}\bar w^{,j]k}({\bm x}_E)-\bar w^{[i,j]k}({\bm x}_E)-\frac12\delta^{k[i}\dot{\bar w}^{,j]}({\bm x}_E)\right]\;,
 \eea
 where the dot over function denotes a total derivative with respect to time, and $v^i_E=dx^i_E/dt$ is the velocity of the geocenter with respect to BCRS.
Notice that the derivatives from the external potentials are calculated on the world line of the geocenter, ${\bm x}_E={\bm x}_E(t)$, so that, for example, $\bar w_{,ij}({\bm x}_E)=\bar w_{,ij}({\bm x})|_{{\bm x}={\bm x}_E}$, etc. 

Vector quantity $Q_i$ is a small acceleration of the geocenter\rq{}s world line with respect to a geodesic world line that is caused by the coupling of the dipole, $I^i_E$, and quadrupole, $I^{ij}_E$, moments of Earth\rq{}s gravitational field to the tidal gravitational field of the external bodies \citep{Kopeikin_2011_book}, 
\be
Q_i=\frac1{M_E}\left(\ddot{I}_E^i-Q_{ij}I^j_E-\frac12Q_{ijk}I_E^{jk}\right)\;.\\
\ee
The quadrupole moment, $C_{ik}$, is a post-Newtonian matrix of the dynamic rotation of the GCRS spatial axes with respect to the BCRS ones which appears in the metric tensor because of the IAU 2000 resolution demanding to keep the spatial axes of the BCRS and GCRS aligned to make both coordinate system kinematically-nonrotating \citep{BK_1989_1, DSX_1}. This demand, however, makes GCRS dynamically rotating with the post-Newtonian angular velocity corresponding to the infinitesimal matrix of rotation $C_{ij}$. With a sufficient accuracy the mathematical expression for the matrix of the post-Newtonian dynamic rotation reads \citep{Kopeikin_2011_book}
\be
C_{ij}=-\bar w^{[i,j]}({\bm x}_E)+\frac34v_E^{[i}\bar w^{,j]}({\bm x}_E)+\frac14v_E^{[i}Q^{j]}\;,\\
 \ee 
where the first term in the right side is the gravitomagnetic (Schiff or Lense-Thirring) precession, the second term is the geodetic (de Sitter) precession, and the third term is the Thomas precession caused by the non-geodesic motion of Earth\rq{}s geocenter with acceleration $Q_i$.

The asymptotic matching technique allows us to derive the transformation law between time and spatial coordinates of BCRS and GCRS \citep{Kopeikin_2011_book}. The post-Newtonian transformation between the spatial coordinates of GCRS and BCRS reads \citep{soffel2003,petit_2010} 
\begin{equation}\label{trans-X}
X^i = r^i + {1 \over c^2} \left[\frac12v_E^i (\vv_E \cdot \vr) + \wbar({\bm x}_E) r^i + r^i (\va_E \cdot \vr) - \frac12a_E^i r^2 \right] + \cO(c^{-4}) \, ,
\end{equation}
where $\vr \equiv \vx - \vx_E(t)$,  and $\va_E$ is the coordinate acceleration of the Earth\rq{}s geocenter, $\va_E =d{\bm v}_E/dt= d^2 \vx_E /dt^2$.

Post-Newtonian transformation of time coordinates, TCG=$T$ and TCB=$t$, is more complicated.  According to the IAU 2000 Resolutions  it reads \citep{soffel2003,petit_2010}
\be\label{time-trans}
T = t - {1 \over c^2} \left[ A(t) + \vv \cdot \vr \right] 
  + {1 \over c^4} [B(t) + B^i(t) r^i + B^{ij}(t) r^i r^j + C(t,\vx)] + \cO(c^{-5}) \, , 
\ee
with
\begin{eqnarray}
\frac{dA(t)}{ dt}  &=& {1 \over 2} \vv_E^2 + \wbar({\bm x}_E)  \, , \\
\frac{dB(t)}{dt} &=& - {1 \over 8} \vv_E^4 - {3 \over 2} \vv_E^2 \wbar({\bm x}_E) + 4 v_E^i \wbar^i({\bm x}_E) + {1 \over 2} \wbar^2({\bm x}_E) \, , \\
B^i (t) &=& - {1 \over 2} \vv_E^2 v_E^i + 4 \wbar^i({\bm x}_E) - 3 v_E^i \wbar({\bm x}_E) \, \\
B^{ij} (t) &=& - v_E^{(i}  Q^{j)} + 2 \wbar^{(i,j)}({\bm x}_E) - v_E^{(i} \wbar^{,j)}({\bm x}_E) +\frac12 \delta^{ij}\dot \wbar({\bm x}_E) \, , \\
C(t,\vx) &=& - {1 \over 10} r^2 (\dot a_E^i r^i) \, .
\end{eqnarray}
Here, the dot stands for the total time derivative with respect to time $t$, {\it i.e.},
\begin{equation}
\dot \wbar \equiv \wbar_{,t} + v_E^i \wbar_{,i} \;,
\end{equation}
and the round parentheses denote the symmetrization with respect to two spatial indices.

\section{International Terrestrial Reference System}
\label{sec:ter-system}

The science of geodesy primarily deals with observations and measurements conducted by terrestrial observers located on Earth. Therefore, it is practically convenient to work in the rotating geocentric coordinate system - the international terrestrial reference system (ITRS) which time coordinate is TCG coordinate time $T$ (the same as the time coordinate in GCRS) and the spatial coordinates are denoted  $X_{\scriptscriptstyle\sf{ITRS}}^i=\lt(X_{\scriptscriptstyle\sf{ITRS}}^1,X_{\scriptscriptstyle\sf{ITRS}}^2,X_{\scriptscriptstyle\sf{ITRS}}^3\ri)$. Transformation from GCRS to ITRS is given by the IERS Conventions \citep{petit_2010,Kopeikin_2011_book}
\be\la{kop6}
X_{\scriptscriptstyle\sf{ITRS}}^i=\Lambda^{ij}X^j\;,
\ee 
where $\Lambda^{ij}\equiv\Lambda^{ij}(T)$ is the orthogonal matrix of rotation depending on time $T$. Due to the property of the orthogonal matrices the inverse matrix $\Lambda^{-1}$ of the transformation coincides with the transpose matrix $\lt(\Lambda^{-1}\ri)^{ij}=\Lambda^{ji}$ so that the inverse transformation between GCRS and ITRS is
\be\la{kop7}
X^i=\Lambda^{ji}X_{\scriptscriptstyle\sf{ITRS}}^j\;.
\ee  

According to the IERS theory of the Earth rotation \citep{petit_2010} the matrix $\Lambda^{ij}$ can be represented in two equivalent forms corresponding to the, so called, CIO-based transformation and equinox-based transformation (see \citep[Chapter 5.9]{petit_2010} and \citep[Equation 9.75]{Kopeikin_2011_book}). For analytic consideration the equinox-based transformation is more convenient and therefore used in the following. The matrix of the equinox-based transformation is represented as a product of four matrices
\be\la{kop8}
\Lambda^{ij}=W^{ki}(T)R_3^{kp}\lt(\rm{GAST}\ri)N^{pq}(T)P^{ql}(T)B^{lj}\;,
\ee
where the time $T$=TCG, $W^{ki}$ is the matrix of the polar wobble, $R_3^{kp}\lt(\rm{GAST}\ri)$ is the matrix of the diurnal rotation depending on the Greenwich Astronomical Sidereal Time (GAST) that is a function of time $T$=TCG, $N^{pq}$ is the matrix of nutation, $P^{ql}$ is the matrix of precession, and the constant matrix $B^{lj}$ describes the, so-called, frame bias.

These matrices have the following structure \citep{petit_2010}
\ba\la{kop9}
P^{ij}&=&R_3^{ik}(\chi_A)R_1^{pk}(-\omega_A)R_3^{qp}(-\psi_A)R_1^{qj}(\varepsilon_0)\;,\\\la{kop10}
N^{ij}&=&R_1^{ki}(-\varepsilon-\Delta\varepsilon)R_3^{pk}(-\Delta\psi)R_1^{pj}(\varepsilon)\;,\\\la{kop11}
W^{ij}&=&R_3^{ki}(-s')R_2^{kp}(x_{\sf p})R_1^{pj}(y_{\sf p})\;,
\ea
where $\varepsilon_0$ and $\varepsilon=\varepsilon(T)$ are correspondingly a constant and instantaneous obliquity of the celestial equator to ecliptic, $\chi_A=\chi_A(T)$, $\omega_A=\omega_A(T)$, $\psi_A=\psi_A(T)$ are secular variations in precession, $\Delta\epsilon=\Delta\epsilon(T)$ and $\Delta\psi=\Delta\psi(T)$ are periodic variations in nutation, $x_{\rm p}=x_{\rm p}(T)$ and $y_{\rm p}=y_{\rm p}(T)$ describe the polar wobble, and $s'=s\rq{}(T)$ is the small secular variation describing the shift between the ITRS origin of longitude and the terrestrial intermediate origin (TIO) \citep{Kopeikin_2011_book, petit_2010}. 

The components of the rotational matrix are used to calculate the GCRS velocity of motion of a terrestrial observer (clocks) located on Earth\rq{}s surface,
\be\la{kop12}
v^i_{\scriptscriptstyle\sf{GCRS}}=\epsilon^{ijk}\omega^jX_{\scriptscriptstyle\sf{ITRS}}^k+\dot X_{\scriptscriptstyle\sf{ITRS}}^i\;,
\ee
where the overdot denotes a time derivative, $\epsilon^{ijk}$ is a fully anti-symmetric symbol of Levi-Civita with $\epsilon^{123}=+1$, $X_{\scriptscriptstyle\sf{ITRS}}^i$ is the ITRS coordinate of the observer (clock), $\dot X_{\scriptscriptstyle\sf{ITRS}}^i$ is the residual velocity of the observer (clock) with respect to ITRS due to various geophysical reasons or simply because the observer is moving in a car or aircraft, and $\omega^i=\omega^i(T)$ is the instantaneous angular velocity of the rotation of the ITRS with respect to GCRS,
\be\la{kop13}
\omega^i=-\epsilon^{ijk}\Lambda^{jp}(T)\frac{d}{dT}\Lambda^{kp}(T)\;.
\ee
In what follows it will be convenient to split the ITRS position of the observer (clock) in two components
\be\la{kop13a}
X_{\scriptscriptstyle\sf{ITRS}}^i=X_{\scriptscriptstyle\sf{geoid}}^i+ \int^H_0n^i(h)dh\;,
\ee
where $n^i(h)$ is a unit vector along the direction of the plumb line passing through the point of observation, $X_{\scriptscriptstyle\sf{ITRS}}^i$  located on Earth surface, $X_{\scriptscriptstyle\sf{geoid}}^i$ is the point on geoid connected by the plumb line to the position of the observer on surface, $dh$ is the element of length along the plumb line, and $H$ is the orthometric height at the observer's position \citep{Torge_2012_book}. 

Precise analytic expression for the angular velocity, $\omega^i$, is too complicated and we develop it only up to a linear approximation with respect to the variations of the parameters entering (\ref{kop9})--(\ref{kop11}). For this purpose we use the following approximations
\ba\la{kop14}
\chi_A&=&\delta\chi_A\;,\\\la{kop15}
\psi_A&=&\delta\psi_A\;,\\\la{kop16}
\omega_A&=&\varepsilon_0+\delta\omega_A\;,\\\la{kop17}
\varepsilon&=&\varepsilon_0+\delta\varepsilon\;\\\la{kop18}
{\rm GAST}&=&T+\Delta\psi\cos\varepsilon_0+\delta T\;,
\ea
where the very last terms in (37)-(41) are functions of time, $T$=TCG, which are changing due to the systematic (both periodic and secular) variations in the spatial orientation of the Earth's rotational axis, wobble, and tidal friction \citep{Kopeikin_2011_book,petit_2010}.

Then, in the linear approximation the components of the angular velocity of Earth\rq{}s rotation are
\begin{align}\la{kop19}
\omega^1=&\dot{x}_{\sf p}+\Omega y_{\sf p}-\sin(\Omega T)\frac{d}{dT}\Bigl[{\delta\omega_A+{\Delta\e}}\Bigr]-\cos(\Omega T)\sin\e_0\frac{d}{dT}\Bigl[{\delta\psi_A}+{\Delta\psi}\Bigr],\\\la{kop20}
\omega^2=&\dot{y}_{\sf p}-\Omega x_{\sf p}+\cos(\Omega T)\frac{d}{dT}\Bigl[{\delta\omega_A+{\Delta\e}}\Bigr]-\sin(\Omega T)\sin\e_0\frac{d}{dT}\Bigl[{\delta\psi_A}+{\Delta\psi}\Bigr],\\\la{kop21}
\omega^3=&\Omega+\frac{d}{dT}\Bigl[\delta T+\delta\chi_A-\delta\psi_A\cos\e_0\Bigr]\,.
\end{align}
where $\Omega$ is a fixed value of the angular velocity of the Earth rotation adopted by IAU \citep{petit_2010}.

\section{Time dilation fundamental equation}
\label{sec:sb_grav_time}

We consider a network of atomic clocks located on Earth's surface at different geographic positions. Each clock moves with respect to the GCRS along world line $X^i\equiv X^i(T)$. According to general relativity each clock measures its own proper time $\tau$ which is defined by the equation $-c^2d\tau^2=ds^2$ where the interval $ds$ must be calculated along the world line of the clock. In terms of the GCRS metric tensor, the interval of the proper time reads
\be\la{kop100}
d\tau^2=-\left(G_{00}+\frac{2}{c}G_{0i}v^i+\frac1{c^2}G_{ij}v^iv^j\right)dT^2\;,
\ee
where $v^i$ is velocity of clock with respect to GCRS, and $T$=TCG. In case of clocks in space $v^i$ is the orbital velocity of the spacecraft carrying the clok. If the clock is located on Earth\rq{}s surface, the GCRS velocity of the clock is given by (\ref{kop12}).

After replacing the GCRS metric (\ref{pnmetric}) in (\ref{kop100}) and extracting the root square, we get the fundamental time delay equation
\be\label{kop4}
\frac{d\tau}{dT}=1-\frac{W}{c^2}+{\cal O}\left(c^{-5}\right)\;,
\ee
where time-dependent function $W=W(T)$ is given by \citep{kopejkin_1991,kopeikin_2016prd}
\be\la{kop5}
W=\frac12v^2+V+\frac1{c^2}\left(\frac18 v^4+\frac32 v^2 V-4v^i V^i-\frac12 V^2\right)\;.
\ee
Function $W$ is the post-Newtonian gravity potential taken at the point of localization of the clock. Notice that it includes both the effects of the gravitational field of the Earth and the external bodies (Sun, Moon, planets) in the form of the tidal terms as shown in equation (\ref{dec321}) of the present paper. The clock's proper time can be calculated by integrating (\ref{kop4}) along the world line of the clock,
\be\la{kop5a}
\tau=\int\limits_{T_{0}}^T\lt[1-\frac{W(T')}{c^2}\ri]dT'+{\cal O}\left(c^{-5}\right)\;,
\ee
where $T_0$ is the initial epoch of the integration.

\section{Implications for the definition of timescales}
\label{section:time}

% {\it Contribution from Gerard} \\

The 13$^{th}$ General Conference on Weights and Measures (CGPM) decided in 
its Resolution 1 in 1967
\footnote{The CGPM reports are available at http://www.bipm.org/fr/worldwide-metrology/cgpm/resolutions.html}
that {\it the second is the duration of 9,192,631,770 
periods of the radiation corresponding to the transition between two hyperfine 
levels of the ground state of the caesium 133 atom}. This definition of the 
time unit called for the adoption of a time scale built by cumulating atomic 
seconds and International Atomic Time TAI was defined in 1970 by the International 
Committee for Weights and Measures as {\it the time reference established by the 
BIH on the basis of the readings of atomic clocks operating in various 
establishments in accordance with the definition of the second}, a definition 
recognized by the $14^{th}$ CGPM in its Resolution 1 in 1971. In 1980 the definition 
of TAI was completed by the Consultative Committee for the Definition of 
the Second, adding {\it TAI is a coordinate time scale defined in a geocentric 
reference frame with the SI second as realized on the rotating geoid as the 
scale unit}. This definition explicitly refers to TAI as a coordinate time, 
hence needing a relativistic approach. In 1988 the responsibility of 
establishing TAI was transferred to the International Bureau for 
Weights and Measures (BIPM) in S\`evres (France).

TAI is not disseminated directly and Coordinated Universal Time UTC, which 
was designed to approximate UT1 (a timescale derived from the rotation of the 
Earth), was chosen as the practical world time reference. Since 1972, UTC differs 
from TAI by an integral number of seconds, changed when necessary by insertion 
of a leap second, as predicted and announced by the International Earth 
Rotation and Reference System Service (IERS). It is not the purpose of 
this paper to discuss this issue; the point is just to remind that changes 
in the definition of timescales may have wide-ranging consequences irrespective 
of the practical implications. 

%{\bf THIS IS TO BE COORDINATED WITH HOW TCB AND TCG HAVE BEEN MENTIONED EARLIER}

As mentioned in previous sections, TCB and TCG are the time coordinates of the BCRS and GCRS, respectively. The 1991 
Recommendation 3 of Resolution A4 of the International Astronomical Union (IAU)
\footnote{The IAU Resolutions are available at http://www.iau.org/administration/resolutions/general\_assemblies/} 
defined the scale unit of TCB and TCG to be consistent with the SI second. This 
means that if readings of the proper time of an observer, expressed in SI seconds, 
are recomputed into TCB or TCG using the formulas from the IAU Resolutions, 
without any additional scaling, one gets corresponding values of TCB or TCG 
in the intended units. The Recommendation also defines the origin of TCB 
and TCG by the following relation to TAI: TCB (resp. TCG) = TAI + 32.184 s 
on 1977 January 1st, 0 h TAI, at the geocenter.

In the following, we assume that the temporal variations in the physical quantities (e.g. potential, equipotential surface) due to tides are taken into account through appropriate reductions so that the quantities are considered as quasi-static, i.e. constant or with a slowly varying secular change. For a clock at rest on the Earth's surface, the relation between proper 
time and coordinate time is given by Eq.~(\ref{kop4}) that is
\begin{eqnarray}
\frac{d \tau}{d\rm TCG} ~= 1 - \frac{W}{c^2}
\label{eq:GP1}
\end{eqnarray}
where for practical purposes, it is sufficient to retain in $W$ only the first two terms of the gravity potential (rotational plus gravitational) given by Eq.~(\ref{kop5}). Any 
time differing from TCG by a constant rate may also be chosen as a coordinate time in 
the geocentric system, and this is the case of Terrestrial Time TT which differs from TCG by a 
constant rate: 
\begin{eqnarray}
\frac{d\rm TT}{d\rm TCG} ~=~ 1- L_G
\label{eq:GP2}
\end{eqnarray}

In the original definition (IAU'1991 Recommendation 4 of Resolution A4), $L_G$ 
is such as {\it the unit of measurement of TT is chosen so that it agrees 
with the SI second on the geoid}, i.e. $L_G = W/c^2$, where $W$ is the latest 
estimate of the gravity potential on the geoid. By IAU Resolution B9 (2000) 
TT is explicitly re-defined with respect to TCG by turning $L_G$ in Eq.~(\ref{eq:GP2}) 
into a defining constant where $L_G ~=~ 6.969290134 \times 10^{-10}$. This constant has 
been computed as $L_G ~=~ W_0/c^2$ taking for $W_0$ the value $62636856$ m$^2$/s$^2$ 
recommended by Special Commission 3 of the International Association of Geodesy (IAG) in the year 1999,  
before the new definition. This value will be noted below $W_0$(2000).

TAI is a realization of coordinate time TT, to within the constant offset 
of 32.184 s, and is subject to uncertainties in the realization. Timescales 
derived from TAI, i.e. UTC and other realizations steered to UTC such as 
GNSS reference times, are also realizations of TT with the same remarks. 
Because TAI is defined with respect to the rotating geoid, the transformation 
from proper time to TAI requires, in principle, the actual value, noted $W_0$, 
of the gravity potential on the geoid. Relation~(\ref{eq:GP1}) transforms to
\begin{eqnarray}
\frac{d \tau}{d\rm TAI} ~=~ 1 - \frac{W - W_0}{c^2}\;,
\label{eq:GP3}
\end{eqnarray}
where $W_0=W({\bm X}_{\scriptscriptstyle{\sf geoid}})$ and ${\bm X}_{\scriptscriptstyle{\sf geoid}}=\{X^i_{\scriptscriptstyle{\sf geoid}}\}$ is the ITRS coordinate on the geoid.
In practice, the potential $W$ is expanded in the Taylor series around its value on the geoid \citep{Torge_2012_book}
\be
W({\bm X}_{\scriptscriptstyle{\sf ITRS}})=W_0- {\bar g}H+{\cal O}(H^2)\;,
\ee
so that Eq.~(\ref{eq:GP3}) is generally applied as
\begin{eqnarray}
\frac{d \tau}{d\rm TAI} ~=~ 1 + \frac{\bar g H}{c^2}
\label{eq:GP4}
\end{eqnarray}
where $H$ is the orthometric height of the clock and $\bar g$ the average value 
of the acceleration of gravity between the geoid and the clock. 

%{\bf REFER TO A PREVIOUS SECTION FOR DETAILS}

Because a change in height of one meter causes a change in rate of about $1 \times 10^{-16}$, 
Eq.~(\ref{eq:GP4}) can be somewhat loosely applied for present-day Caesium primary frequency 
standards which have accuracy of order $1 \times 10^{-16}$, assuming that the geoid, the reference level of the height system and the 
height are all correctly realized or measured with an uncertainty well below one metre. 
This is no more the case when considering a clock accuracy of the order $1 \times 10^{-17}$ 
and below for which an uncertainty of order 1 cm is needed. For better accuracy, 
with validity about $5 \times 10^{-19}$, Eq.~(\ref{eq:GP1}) must include tidal terms $V_{\sf tide}\equiv\bar V$ given in (\ref{tidepot1}) to obtain
\begin{eqnarray}
\frac{d \tau}{d \rm TCG} ~=~ 1 - \frac{1}{c^2} \left(\frac12 v^2+V_E + V_{\sf tide}\right)
\label{eq:GP5}
\end{eqnarray}
so that
\begin{eqnarray}
\frac{d \tau}{d\rm TT} ~=~ 1 +\frac{1}{c^2}\left(gH+V_{\sf tide}\right)\;.
\label{eq:GP6}
\end{eqnarray}
The tidal potential has been considered in full detail, for example, in \citep{Agnew2007}. In the quadrupole gravitational field approximations it reads $V_{\rm tide}\equiv V_2$, i.e., we take into account  in (\ref{tidepot1}) merely the term depending on the external quadrupole moment $Q_{ij}$. This expression is well known in scientific literature (see, for example, \citep{Agnew2007,Torge_2012_book,Simon_2013}), and can be presented in the form
\be\la{kop30}
V_2=V_2^{\rm zonal}+V_2^{\rm tesseral}+V_2^{\rm sectorial}\;,
\ee
where
\ba\la{kop31}
V_2^{\rm zonal}&=&\sum_{A\not=E}{\cal D}_A\lt(\frac{\bar r_{\scriptscriptstyle{\sf EA}}}{r_{\scriptscriptstyle{\sf EA}}}\ri)^3\lt(\frac13-\sin^2\varphi\ri)\lt(1-3\sin^2\delta_A\ri)\\\la{kop32}
V_2^{\rm tesseral}&=&\sum_{A\not=E}{\cal D}_A\lt(\frac{\bar r_{\scriptscriptstyle{\sf EA}}}{r_{\scriptscriptstyle{\sf EA}}}\ri)^3\sin 2\varphi\sin 2\delta_A\cos h_A\\\la{kop33}
V_2^{\rm sectorial}&=&\sum_{A\not=E}{\cal D}_{A}\lt(\frac{\bar r_{\scriptscriptstyle{\sf EA}}}{r_{\scriptscriptstyle{\sf EA}}}\ri)^3\cos^2\varphi\cos^2\delta_A\cos 2h_A\;,
\ea
where 
\be
{\cal D}_A=\frac34 GM_A\frac{r^2}{\bar r_{\scriptscriptstyle{\sf EA}}^3}\;,
\ee
is the Doodson's constant for the body A, $\bar r_{\scriptscriptstyle{\sf EA}}$ is the reference (mean) value of the distance $r_{\scriptscriptstyle{\sf EA}}$ during one revolution of the body A around Earth, $\delta_A$, $h_A$ are respectively the declination and the hour angle of the external body A in the ITRS, and the summation goes over all the external bodies of the solar system but Earth.

Note that all developments in this section are valid to an accuracy level of order $1 \times 10^{-18}$ and should be re-examined to reach an accuracy level 
of order $10^{-19}$ ($\equiv$ mm in height). Then, e.g., the full post-Newtonian gravity potential given in Eq.~(\ref{kop5})  as well as the geoid in its relativistic definition should be considered, \cf~\citep{Kopeikin_2011_book}.

Consider the transformation to TCG (the case of TAI will be considered further below) with Eq.~(\ref{eq:GP5}), one needs to determine the gravity 
potential at the location of the clocks. This used to be a difficult task at the $10^{-17}$ 
level worldwide \citep{petit1997}, however recent global satellite geopotential 
models allow this determination at the level of a few parts in $10^{18}$ \citep{denker2013}. 
On the other hand 
the centrifugal potential as well as other time-varying tidal effects due 
to external masses (direct effect and indirect effect from Earth and ocean 
tides) can reach a few $10^{-17}$ but can be computed at the $1 \times 10^{-18}$ 
accuracy level \citep{voigt2016}. In addition, other non-tidal effects due to mass redistributions 
in the Earth, the hydrosphere or the atmosphere are to be considered at the 
$10^{-18}$ accuracy level. It is not the purpose of this paper to study 
these effects but, contrary to tidal effects, it is not straightforward 
to model them and to define a conventional reference for these effects.
Such $10^{-18}$ accuracy in the computation of the relativistic frequency shift will be necessary for the realization of timescales 
when new clocks with such frequency accuracy will provide the definition of the second.

Considering that the best Caesium primary frequency standards have an accuracy of order $1-2 \times 10^{-16}$, the present requirements are less stringent.
In past years, it was claimed an uncertainty for the relativistic frequency shift of 
$2-3 \times 10^{-17}$ for the location of the NIST in Boulder (Colorado, USA) 
\citep{pavlis2003} and of $1 \times 10^{-17}$ for the INRIM in Torino (Italy) 
\citep{calonico2007}. A review of the frequency standards contributing to TAI in 2015  
\footnote{BIPM report on Time activities available at http://www.bipm.org/en/bipm/tai/annual-report.html}
shows that a good fraction adopts a conservative approach in evaluating the relativistic frequency shift e.g. $1 \times 10^{-16}$ is used for 
the LNE-SYRTE in Paris (France) and for the NIM in Beijing (China), 
$0.5 \times 10^{-16}$ for the NPL in Teddington (UK) and for the VNIIFTRI 
in Mendeleevo (Russia), while the NIST adopts $0.3 \times 10^{-16}$ as mentioned 
above. Some metrological centers, however, like the INRIM or the PTB in Braunschweig (Germany) 
report an uncertainty at or slightly below $0.1 \times 10^{-16}$.

The geoid appearing in the definition of TAI is classically defined as the 
level surface of the gravity potential closest to the topographic mean 
sea level. Therefore the value of the potential on the geoid $W_0$ depends 
on the global ocean level which changes with time due to geophysical reasons. Several authors have 
considered the time variation of $W_0$, see e.g. \citep{bursa2007,dayoub2012}, 
but there is some uncertainty in what is accounted for in such a linear model. 
A recent estimate \citep{dayoub2012} over 1993-2009 is 
$d W_0 / dt = -2.7 \times 10^{-2} ~ \rm m^2 s^{-2}yr^{-1}$, mostly driven by the 
sea level change of +2.9 mm/yr. The rate of change of the global ocean level 
could vary during the next decades, nevertheless, to state an order of 
magnitude, considering a systematic variation in the sea level of 
order 3 mm/yr, different definitions of a reference surface for the 
gravity potential could yield differences in frequency of order 
$3 \times 10^{-18}$ in a decade. In addition, there are numerous 
questions associated with realizing a global reference {\it closest to 
the mean sea level}, see \eg\ a discussion in \citep{sanchez2012}, so that 
work is still under way in IAG working groups to provide conventional 
definitions and procedures of realization for a Vertical Datum Standardization.

Therefore, there is evidence of inconsistency in the present set of concepts and/or definitions as summarized hereafter.
On the geodetic side, there is inconsistency between the three propositions:
\begin{itemize}
\item The geoid is the equipotential surface closest to the mean sea level;
\item The value of the gravity potential on the geoid is $W_0$;
\item $W_0$ is a defining (fixed) constant.
\end{itemize}
On the time metrology side, there is inconsistency between the three propositions:
\begin{itemize}
\item TT is a coordinate time defined by IAU Recommendation B9 (2000);
\item TAI is a coordinate time with the SI second as realized on the rotating 
      geoid as the scale unit;
\item TAI is a realization of TT.
\end{itemize}
The inconsistency is at a level of $10^{-17}$ (equivalent to 9 cm in height) 
when considering that such fundamental concepts should be valid for decades. 

The choice of the value $W_0$(2000) provides a formal definition of a 
surface where clocks run at the same rate as TT, which has been named 
the chronometric geoid \citep{wolf1995}. As the {\it classical geoid}
remains linked to the mean sea level, these two surfaces differ and the 
difference evolves with time. One possible solution on the geodetic side 
is to update the $W_0$ value as necessary to reflect the actual mean sea 
level as well as improvements in the determination of $W_0$, so that $W_0$ 
is no more used as a constant but is varying with time. On the time metrology side, 
one solution is to explicitly define TAI as a realization of TT so that 
it is no more linked to the time-varying geoid. Indeed, if the adopted 
value of the gravity potential on the geoid $W_0$ is updated, this has 
no implication on the $L_G$ value which remains a fixed conventional value 
relating TCG and TT. The $W_0$ value and the definition of the geoid 
are the responsibility of the IAG which addressed the issue in its 
Resolution 1 (2015) {\it for the definition and realization of an International 
Height Reference System (IHRS)}. As mentioned in \citep{ihde2015}, {\it For a 
global height reference system, any value $W_0$ within a range of a few decimeters 
can be defined as conventional without affecting the task of defining and 
realizing a global height reference system}. Nevertheless, IAG Resolution 1 
(2015) recommends both a new value of the gravity potential at the geoid $W_0 
= 62636853.4~\rm m^2s^{-2}$, here noted $W_0$(2015), and to use this value 
a reference for the IHRS. Should such IHRS-compatible geopotential values 
become available for vertical reference points, the difference between 
$W_0$(2015) and $W_0$(2000), equivalent to $2.9 \times 10^{-17}$ in rate 
shift, has to be noted when realizing TT from an atomic clock. The 
reformulation of the TAI definition as well as its procedure of 
realization is under the responsibility of the Consultative Committee for Time and Frequency (CCTF) which has formed 
a task group to propose, at its 2017 meeting, a new definition of TAI 
to be adopted by the CGPM in 2018. Changing the wording in the definition 
of TAI will be one task but it may be the easy one. Applying Eq.~(\ref{eq:GP6}) 
will still be a difficult task on a global scale at the $10^{-18}$ level. E.g. while height networks can be directly 
linked when on the same continent, larger uncertainties persist in the 
global determination of the geoid and of vertical references from effects 
affecting the sea surface topography \citep{fu2010}.

Placing ultra-accurate clocks on Earth is inconvenient, because of the large 
number of small geophysical effects that have to be taken into account to compute the 
relativistic frequency shift. However it is also an opportunity because their accuracy gives access to study these effects. As 
has long been realized and dreamed of, it might be an optimal solution to 
have some ultra-accurate clocks in space, to be used as a reference to 
generate TAI and to which Earth-based clocks could be compared. 
Comparisons of ultra-accurate clocks could therefore help in the future 
to establish a worldwide vertical datum.

\section{Height systems and terrestrial gravimetry with clocks}
\label{section:terrestrialy}

Here, we discuss a novel method of the direct use of clock measurements to derive physical heights and to resolve discrepancies in classical height systems. We address time-variable effects that have to be considered when using clock data in practice. These effects include both temporal variations of the gravity field and instabilities in the rotation of the Earth. A particular application is the direct use of observed potential differences for the improved recovery of the regional gravity field in the Massif Central region in the middle of southern France.

\subsection{Height systems and clock measurements}
\label{subsec:heightsys}

% {\it Contribution from J\"urgen}

\begin{figure}[htb]
\begin{center}
\includegraphics[width=1.0\textwidth,angle=0]{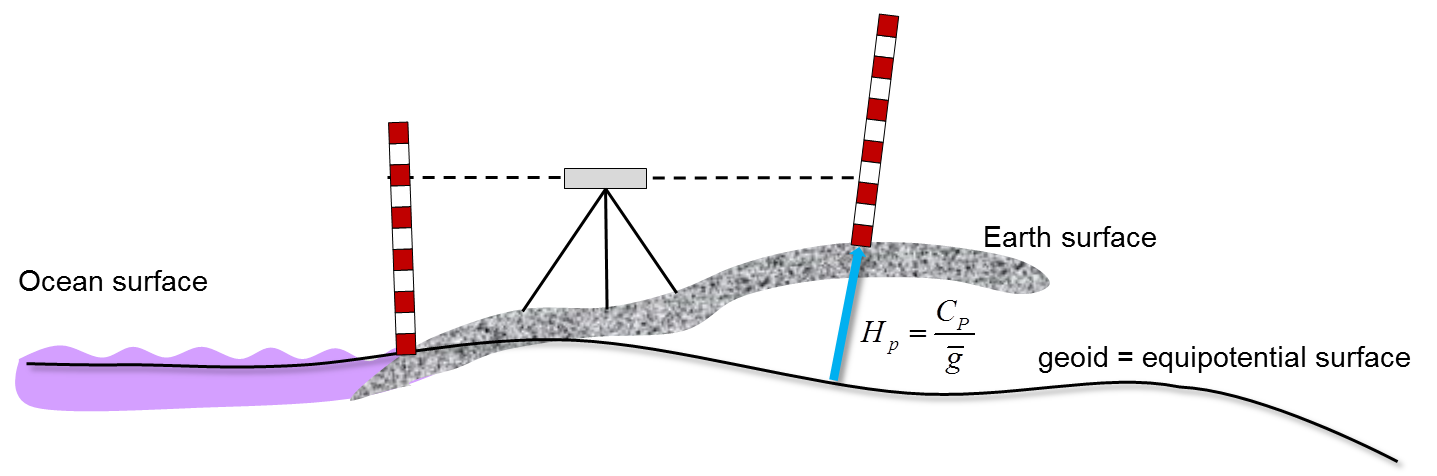}
\end{center}
\caption{Physical heights and geoid}
\label{fig:height-geoid}
\end{figure}

To the major objectives of geodesy belong the determination 
of physical heights (i.e. those heights related to the gravity 
field) and the determination of the corresponding height reference surface (i.e. the geoid or quasi-geoid) \citep{Torge_2012_book},
see Figure~\ref{fig:height-geoid}. For example, the orthometric 
height is defined by
\begin{eqnarray}
H ~=~ \frac{C_P}{\bar g}
\label{ortho-height}
\end{eqnarray}
where $C_P$ is the so-called geopotential number which is the 
difference of the gravity potential between the geoid 
and the surface point $P$, and $\bar g$ is  mean 
gravity along the (curved) plumbline. A similar relation holds for normal 
height $H_N$ (where the geopotential number $C_P$ has to be 
divided by a mean normal gravity value $\bar\gamma$) and 
quasi-geoid. Geopotential numbers $C_P$ are classically 
obtained via geometrical levelling and terrestrial gravimetry. The 
latter is required to correct the raw levelling results to 
consider the non-parallelism of the equipotential surfaces of 
which the geoid is just that one surface corresponding to the 
mean ocean surface at rest. This classical approach has some 
drawbacks such as increasing errors with the length of the 
levelling loops or systematic errors by combining data from 
different measurement periods. This method is very time consuming, 
especially if large areas have to be covered. For detecting 
vertical deformations, repeated levelling measurements are required 
which further increases the mentioned complexity.

When establishing a physical height system, normally one refers 
to a datum point like a selected tide gauge, e.g. NAP (Normaal
 Amsterdams Peil), and then only differences of potential 
numbers $\Delta C_P$ are used further on. The basic network of 
physical heights for a certain area, like a country, is then 
computed at the level of geopotential numbers before they are 
converted into physical heights by dividing the $C_P$ values 
by $\bar g$ or $\bar\gamma$.

If the geoid (or quasi-geoid) shall be determined point-wisely, 
one can use the physical heights together with ellipsoidal GNSS 
(Global Navigation Satellite System) heights
\begin{eqnarray}
N ~=~ h - H,
\label{eq:geoid-height1}
\end{eqnarray}
where $N$ is the geoid height above the ellipsoid, $h$ is the ellipsoidal 
height of the surface point $P$. Similar relations hold for the quasi-geoid 
height: $\zeta = h - H_N$. Alternatively, one could apply the Bruns formula
\begin{eqnarray}
N ~=~ \frac{T_0}{\gamma_0},
\label{eq:geoid-height2}
\end{eqnarray}
where the geoid height $N$ is derived from the disturbing potential 
$T_0$ at the geoid and the normal gravity value $\gamma_0$ taken 
at the ellipsoid. A similar expression holds again for the quasi-geoid 
height $\zeta = T_P/\gamma_Q$, now the disturbing potential is needed 
at the surface point $P$ and the normal gravity value at the telluroid 
point $Q$\footnote{The telluroid is an approximation to the Earth 
surface that is consistent to the theory of Molodensky \citep{HeiskanenMoritz_1967}.}. The 
disturbing potential $T_P$ is defined as difference between the gravity potential 
of the Earth $W_P$ and the normal potential $U_P$ which 
is computed from the values of a mean rotational ellipsoid approximating 
the Earth: $T_P = W_P-U_P$, \cf\ \citep{Torge_2012_book}. The disturbing potential can be computed by solving a boundary-value 
problem where various variants exist. Often the long-wavelength 
contributions of the gravity field are taken from satellite solutions 
that are combined with terrestrial gravity and topographic data 
for the shorter wavelengths, \cf\ \citep{denker2013}.
\begin{figure}[htb]
\begin{center}
\includegraphics[width=1.0\textwidth,angle=0]{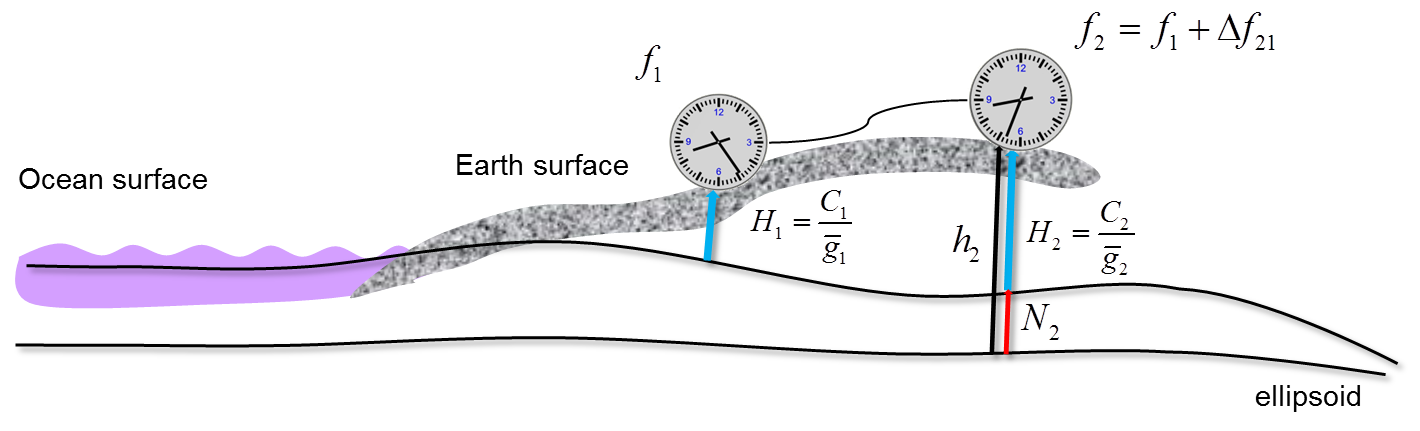}
\end{center}
\caption{Physical heights and clock measurements}
\label{fig:clock-geoid}
\end{figure}

As briefly re-called from geodesy textbooks, the gravity potential 
and mostly differences of the gravity potential are the central 
quantities to derive physical heights and the geoid. In Section \ref{sec:sb_grav_time}
, the dependency of proper time to the gravity potential $W$ is given. Now we consider two clocks with proper times $\tau_1$ and $\tau_2$ (see also Figure \ref{fig:clock-geoid}) which are directly related to their proper frequencies $f_1$ and $f_2$ via 
\begin{eqnarray}
\frac{f_{2}}{f_{1}}=\frac{\tau_{1}}{\tau_{2}}.
\label{eq:df-dtau}
\end{eqnarray}
The corresponding proper frequency difference $\Delta f$ between the two clocks placed at two different locations 
is now derived by introducing Eq.~(\ref{kop4}) in Eq.~(\ref{eq:df-dtau})
\begin{eqnarray}
\frac{\Delta f_{21}}{f_1} ~=~ \frac{\Delta W_{21}}{c^2}.
\label{eq:df-ratio}
\end{eqnarray}
It is directly related to the difference of the gravity potential $\Delta W_{21}$ 
between these locations which again corresponds to the difference 
of the geopotential numbers $\Delta C_{21} = -\Delta W_{21}$. Thus, in future, those 
clock-based gravity potential values can straightforwardly be 
used in physical geodesy. For example, the difference between two orthometric heights can be obtained as (see also \citep{mueller2016})
\begin{eqnarray}
\Delta H_{21} = H_2 - H_1 ~=~ H_1 \frac{\Delta \bar g_{12}}{\bar g_2} + \frac{\Delta C_{21}}{\bar g_2}
\label{eq:dH-otrtho}
\end{eqnarray}
with the difference of the mean gravity values along the plumbline at both locations $\Delta \bar g_{12} = \bar g_1 - \bar g_2$ and the difference of the geopotential numbers $\Delta C_{21} = C_2 - C_1 = W_1 - W_2$, which can now be replaced by clock-based frequency differences $\Delta f_{21}$ according to Eq.~(\ref{eq:df-ratio}):
\begin{eqnarray}
\Delta H_{21}  ~=~ H_1 \frac{\Delta \bar g_{12}}{\bar g_2} - \frac{c^2}{\bar g_2}\frac{\Delta f_{21}}{f_1}
\label{eq:dH-ortho2}
\end{eqnarray}

\subsection{Time-variable effects on clock measurements}
\label{subsec:tempclock}

In the following, we want to discuss some properties that have 
to be considered when using clock measurements and we will give 
a few examples for potential applications where clock data might 
especially be beneficial.

At the accuracy level of mm to cm, it is not sufficient anymore 
to consider gravity potential values as constant in time. They 
are affected by deformations (up to 40 cm due to tidal effects) 
and mass variations. In the following, we always indicate the combined 
effect for both contributions (deformation plus mass effect) as sensed by the clocks. And to make the effect more illustrative, the 
gravity potential value is converted into a 
height value by dividing it by some mean gravity value of about 
9.81 m/s$^2$. 
Even if only potential differences are considered, there are remarkable 
effects of up to 8 cm between PTB in Braunschweig and NPL in London 
caused by relative perturbations of the solid Earth tides 
(\citep{voigt2015,voigt2016}). Those variations with changing amplitudes 
have periods of 12 hours and longer. In addition, many smaller 
effects have to be considered: periodic effects due to ocean 
tides (a few cm), non-tidal oceanic and atmospheric effects 
(1 cm) or episodic effects, e.g., caused by storm surges or 
droughts (1 cm).  Further effects comprise pole tides, variations 
due to land hydrology, tectonic and GIA-induced processes, etc., 
see \citep{voigt2016}. Another class of time-variable effects 
on the gravity potential are variations in Earth orientation.

\vspace{0.5cm}
% {\it Contribution from Sergei}
The Earth orientation effects affect the rate of clocks located on the Earth surface as they are produced by the time-dependent changes in the velocity-dependent potential of the centrifugal force \citep{fateev_2015}. The centrifugal force potential consists of three terms as a direct consequence of the velocity decomposition equation (\ref{kop12})
\be\la{kop21a}
\frac12{\bm v}^2=\frac12{\bm v}_{\scriptscriptstyle{\sf GCRS}}^2=\frac12\lt({\bm\omega}\times{\bm X}{\scriptscriptstyle{\sf ITRS}}\ri)^2+\lt({\bm \omega}\times{\bm X}_{\scriptscriptstyle{\sf ITRS}}\ri)\cdot\dot{\bm X}_{\scriptscriptstyle{\sf ITRS}}+\frac12\dot {\bm X}_{\scriptscriptstyle{\sf ITRS}}^2\;,
\ee
The first term on the right hand side of (\ref{kop21a}) is the centrifugal potential caused by the rotation of ITRS in space. We shall denote it as 
\be\la{kop22}
Z\equiv\frac12\lt({\bm\omega}\times{\bm X}_{\scriptscriptstyle{\sf ITRS}}\ri)^2\;,
\ee
and decompose it in several components corresponding to secular changes in the rotational velocity, precesssion, nutation, and the polar wobble. It is convenient to write down the decomposition in terms of geocentric spherical coordinates associated with ITRS,
\be\la{kop23}
X_{\scriptscriptstyle{\sf ITRS}}^1=r\cos\varphi\cos\lambda\;,\qquad X_{\scriptscriptstyle{\sf ITRS}}^2=r\cos\varphi\sin\lambda\;,\qquad X_{\scriptscriptstyle{\sf ITRS}}^3=r\sin\varphi\;.
\ee
The centrifugal potential can be represented as a superposition
\be\la{kop24}
Z=Z_\oplus+Z_{\rm sec}+Z_{\rm prec}+Z_{\rm nut}+Z_{\rm wob}\;,
\ee
where
\ba\la{kop25}
Z_\oplus&=&\frac12 \Omega^2r^2\cos^2\varphi\;,\\\la{kop26}
Z_{\rm sec}&=&\Omega r^2\cos^2\varphi\frac{d}{dt}\Bigl[\delta T+\delta\chi_A-\delta\psi_A\cos\e_0\Bigr]\;,\\\la{kop27}
Z_{\rm prec}&=&\frac12\Omega r^2\sin 2\varphi\lt[\cos(\Omega t-\lambda)\sin\e_0\frac{d\delta\psi_A}{dt}+\sin(\Omega t-\lambda)\frac{d\delta\omega_A}{dt}\ri]\;,\\\la{kop28}
Z_{\rm nut}&=&\frac12\Omega r^2\sin 2\varphi\lt[\cos(\Omega t-\lambda)\sin\e_0\frac{d\Delta\psi}{dt}+\sin(\Omega t-\lambda)\frac{d\Delta\epsilon}{dt}\ri]\;,\\\la{kop29}
Z_{\rm wob}&=&\frac12\Omega^2 r^2\sin 2\varphi\lt(x_p\sin\lambda+y_p\cos\lambda\ri)
\ea

The second term in (\ref{kop21a}) describes the coupling of the rotational velocity of ITRS and the residual velocity $\dot{\bm X}{\scriptscriptstyle{\sf ITRS}}$ of clocks due to their motion with respect to ITRS. Taking the time derivative from the ITRS coordinates (\ref{kop23}) of clocks and calculating, we get
\be\la{kop29a}
Z_{\rm c}\equiv\lt({\bm \omega}\times{\bm X}_{\scriptscriptstyle{\sf ITRS}}\ri)\cdot\dot{\bm X}{\scriptscriptstyle{\sf ITRS}}=\Omega r^2\cos^2\phi\dot\lambda=2Z_\oplus\frac{\dot\lambda}{\Omega}\;,
\ee
We can observe that the vertical speed of clocks and the motion of the clocks along meridian does not contribute to $Z_{\rm c}$, only the longitudinal component of the velocity matters. The term $Z_{\rm c}$ exceeds $10^{-18}$ if the tangential residual velocity of clocks on the Earth's equator exceeds 0.2 mm/s. We shall assume that clocks are at rest which allows us to neglect the contribution of $Z_{\rm c}$ to the velocity-dependent potential along with
the third term in (\ref{kop21a}). In important cases of clock transportations on cars or aircrafts the term $Z_c$ must be retained.
All rotational effects can well be modelled. In future, some of 
those time-variable effects might even be measured in clock networks.

\subsection{Clocks for datum connection and uncertainties in height systems}
\label{subsec:datum}

What is the major benefit of highly-precise clock measurements? 
Clocks can connect distant areas, i.e. they deliver physical height 
differences for the observed points without being affected by 
levelling errors or by some smoothing effect when combined global 
gravity field models, e.g., from GOCE are used. No (good) ground 
gravity data from terrestrial gravimetry is needed to fill-in 
regional to local gravity variations. This might especially be 
interesting in (relatively) unsurveyed countries where no gravity data 
are available at all, where terrestrial data are difficult to 
obtain (e.g. in rain forests) or in areas with rough environment 
(e.g. in mountain areas).

Clock measurements can be used as independent observations to 
resolve discrepancies in (classical) realisations of height 
systems and geoid solutions, see example in the Alps 
\citep{denker2015}. Even in Germany between north and south over only 700 km, there is a 4 cm discrepancy when determining the quasi-geoid difference by different methods, such as GNSS/levelling versus gravimetric solutions. A similar comparison between NPL, London and PTB, Braunschweig showed differences between the classical methods of up to 15 cm  \citep{denker2016}. These differences are most probably caused by geometric levelling that accumulates measurement errors with increasing distance.

\begin{figure}[htb]
\begin{center}
\includegraphics[width=1.0\textwidth,angle=0]{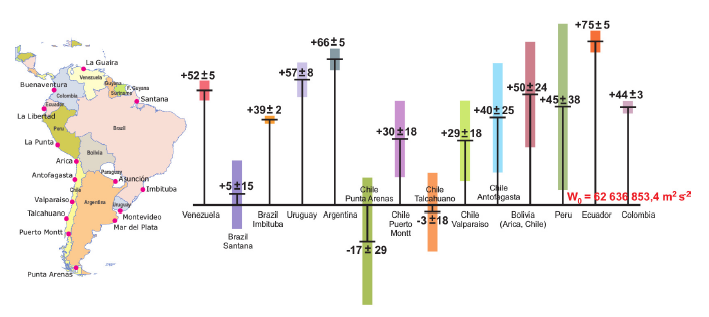}
\end{center}
\caption{Difference of regional vertical reference levels to a global 
one (related to $W_0$) in South America derived from geodetic 
measurements (gravity, GNSS, levelling, satellite altimetry), 
unit: cm (figure is taken from \citep{sanchez2015}}
\label{fig:sirgas-heights}
\end{figure}

Nevertheless in Europe, the connection of national height systems (i.e. of the specific vertical datums of each country) might be 
possible by applying  standard geodetic methods - with the above mentioned uncertainties. But in regions 
like South America, large differences with large error bars are 
present when comparing various tide gauges, especially in the 
southern part (see Figure~\ref{fig:sirgas-heights} taken from \citep{sanchez2015})
where differences in the decimeter level are present. Here, 
immediate improvements could be achieved, even with recent clock 
technology at the few cm level of accuracy, assuming that these 
clocks could be placed at those tide gauges and be connected by 
dedicated fiber links. Besides just taking raw difference 
measurements, one could carry out a new adjustment for the height 
differences where clock, GNSS and altimetry data are used as 
joined input.

Another application could be running clocks on ships to independently 
control altimetric measurements. The ocean surface roughly approximates 
an equipotential surface in the Earth gravity field (then clock rates 
would not vary), but deviate from it because of ocean currents like 
the gulf stream. These deviations reach up to 1 to 2 m and are observed 
by satellite altimetry together with the marine geoid, e.g., derived 
from GOCE data \citep{pail2010}. One idea now is to use combined clock 
and GNSS measurements along selected profiles that can be compared to 
the results of altimetric measurements. Once again, possible systematics 
might by revealed by the clock measurements. Running a clock on a 
moving platform without interruption, however, is another challenge 
that is not discussed here. And, it might even be more challenging when the moving clock has to 
be (frequently) compared to a remote clocks with the same accuracy.
%
% Should we give the formulas for a moving clock, e.g. from Nelson 
% (2011, Metrologia)???
%
% Somewhere the contribution of Isabelle should now be included with 
% the discussion of a best distribution of a clock network for geoid 
% computation.
%
% Should we also mention the relativistic definiton of the geoid? 
% Should we mention GGOS?
% Should we mentions the potential of clocks as new vertical datum, 
% reference points? Frequency as height reference? ...
%

\subsection{Clocks for high resolution geopotential recovery}
\label{subsec:highres_ground}

\begin{figure}
\centering
\begin{tabular}{ccc} 
\includegraphics[width=0.30\textwidth]{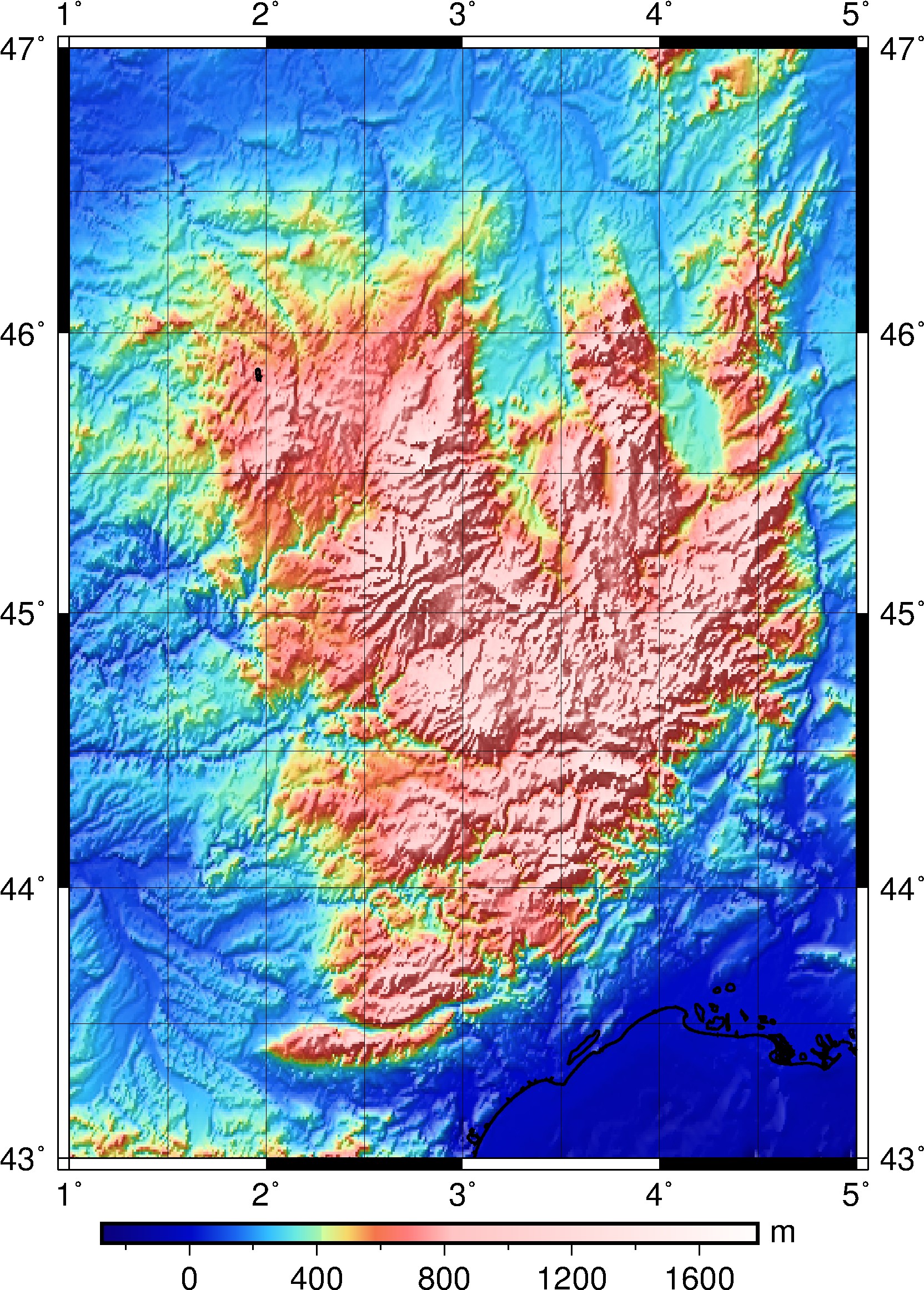} &
\includegraphics[width=0.30\textwidth]{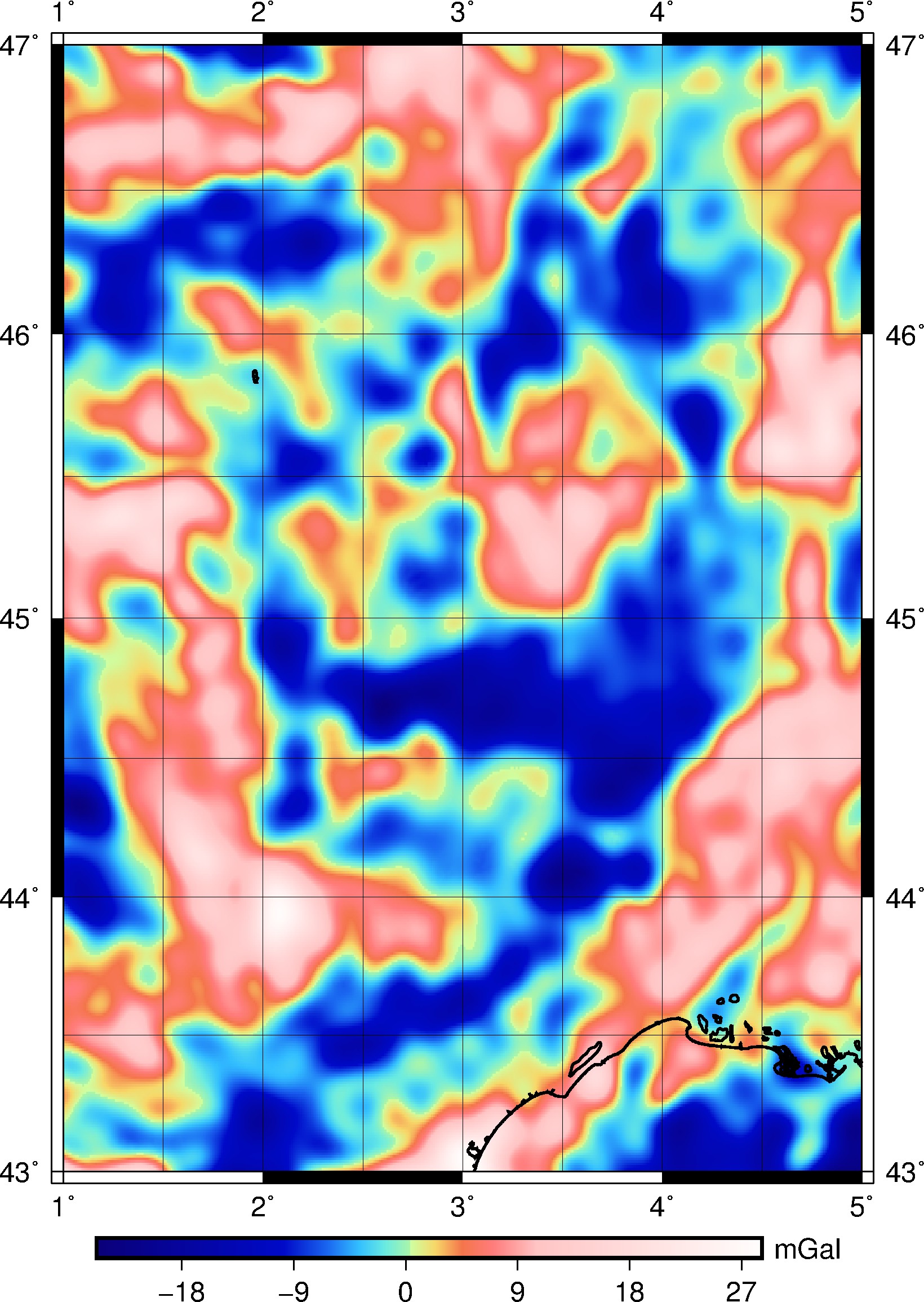}  &
\includegraphics[width=0.30\textwidth]{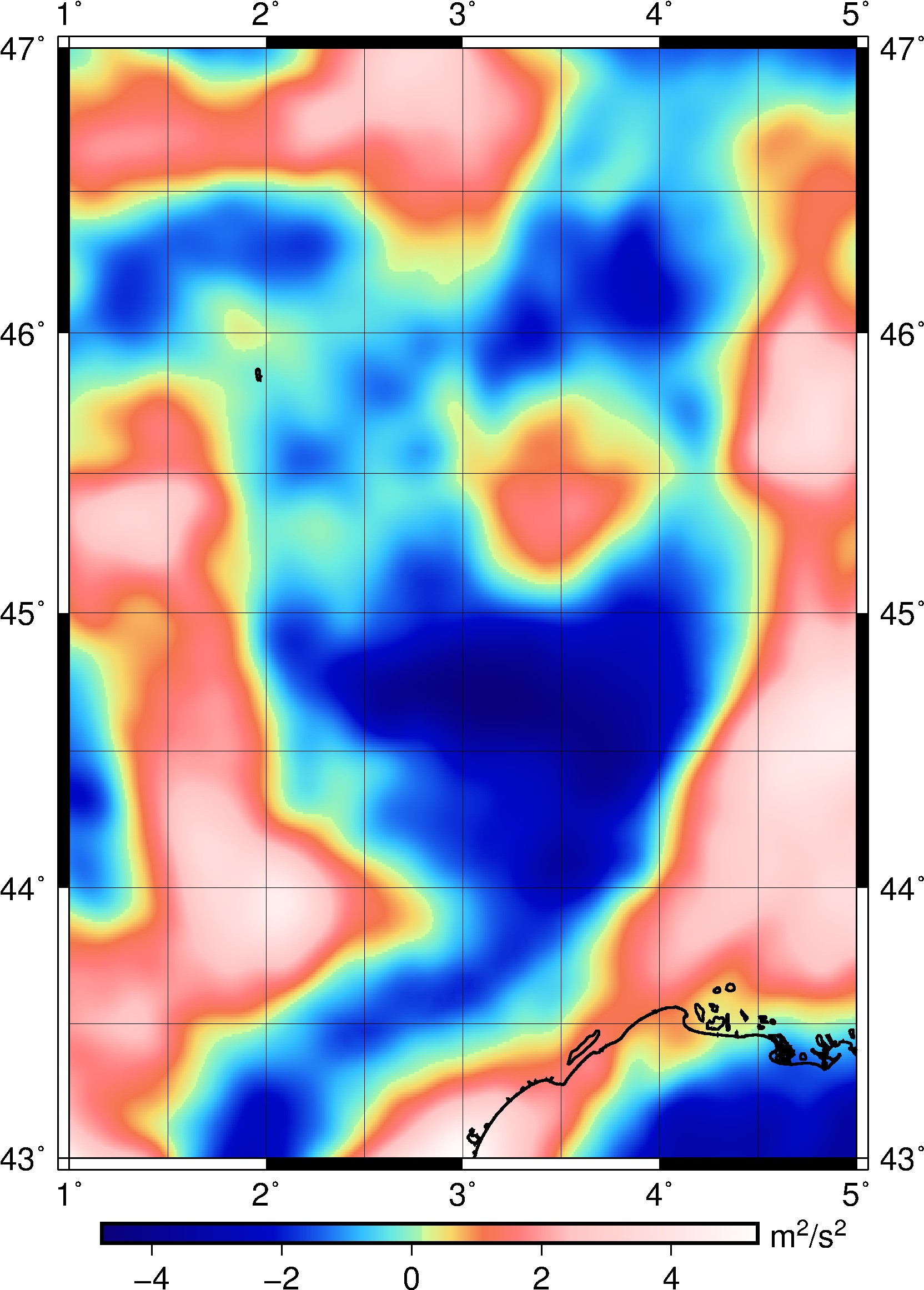} \\
(a) Topography & (b)  $\Xref{\grdist}$ & (c) $\Xref{T}$ \\
\end{tabular}
\caption{Topography and reference grids of the synthetic 
         field~$\Xref{\grdist}$ and~$\Xref{T}$ in Massif Central area. 
         Anomalies are computed at the Earth's surface from the {EIGEN-6C4} 
         model up to d/o 2000. Topography is obtained from the 30~m 
         digital elevation model over France by IGN (Institut National de l'Information G\'{e}ographique et Foresti\`{e}re), completed with \citep{Smith_1997aa} bathymetry and SRTM (Shuttle Radar Topography Mission) data.}
\label{fig:Auv}
\end{figure}

To illustrate what could be the benefit of clocks for high-resolution geopotential 
recovery, beyond the resolution of the satellites and as complement to near-surface 
gravity data, we perform simple simulations where we compare the quality of the 
geopotential reconstruction with  and without adding clock data. We consider the 
Massif Central area, marked by smooth, moderate altitude mountains and volcanic plateaus, 
see Figure~\ref{fig:Auv} (left), which leads to variations of the gravitational 
field over a range of spatial scales, as illustrated in Figure~\ref{fig:Auv} 
(middle and right).

For the tests, we proceed as follows. We sample synthetic gravity and potential 
data from a spherical harmonics geopotential model, and we build a geopotential 
control grid. We estimate the gravity potential by least-squares collocation from 
the synthetic data. We compare the recovered potential on the control grid, to 
the reference one to assess the improvements that can be obtained using clock
data. In this process, different parameters of the problem can be varied, as 
the spatial density of clock data for instance, studied hereafter.

\begin{figure}[!htb]
\centering
\begin{tabular}{p{0.9\textwidth}}
	\hspace{-1.7cm}
\includegraphics[width=1.28\linewidth]{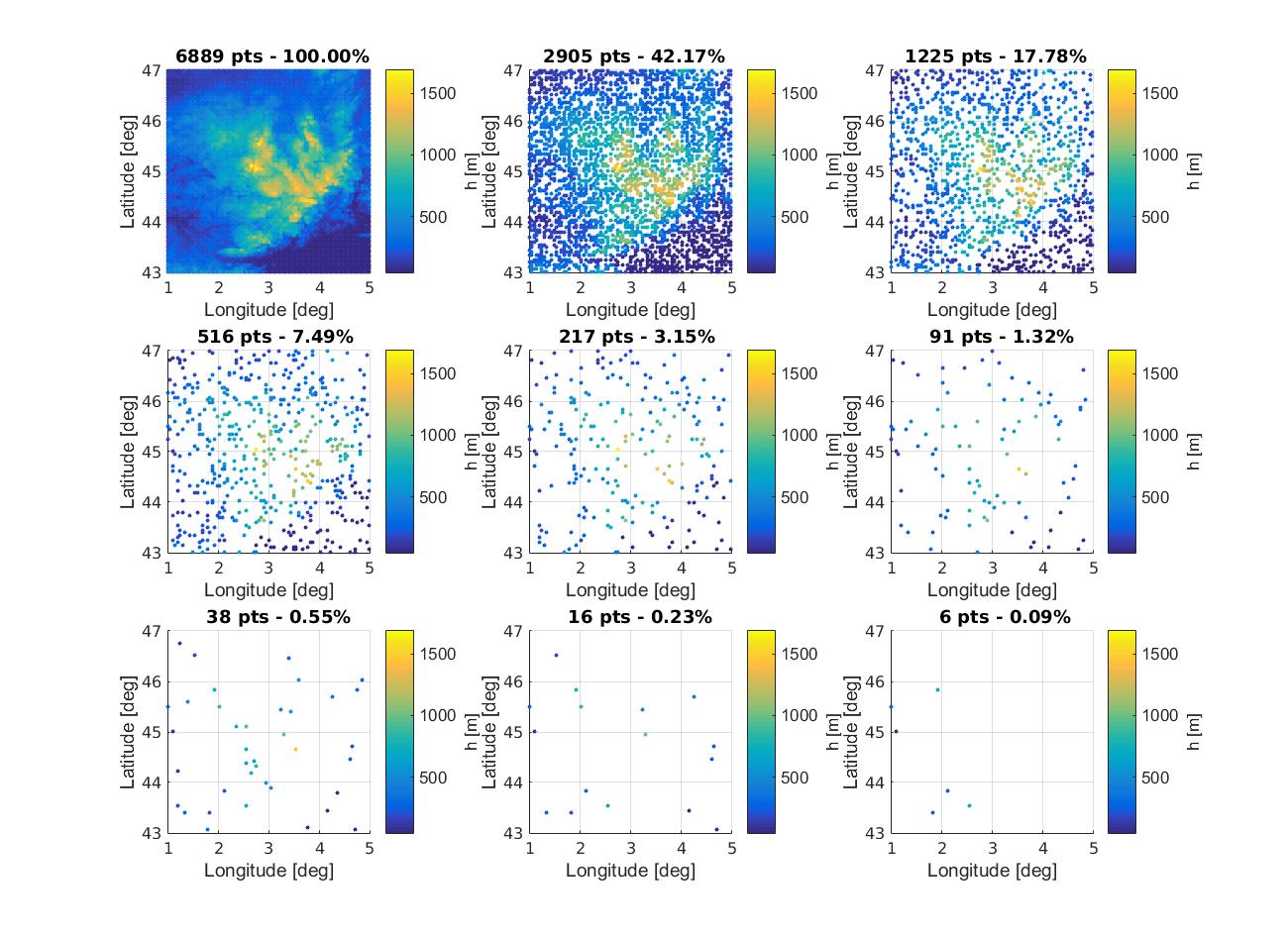} \\
(a) Clock coverages randomly generated from a regular 5-km step grid. \\
\multicolumn{1}{c}
{\includegraphics[width=0.7\linewidth]{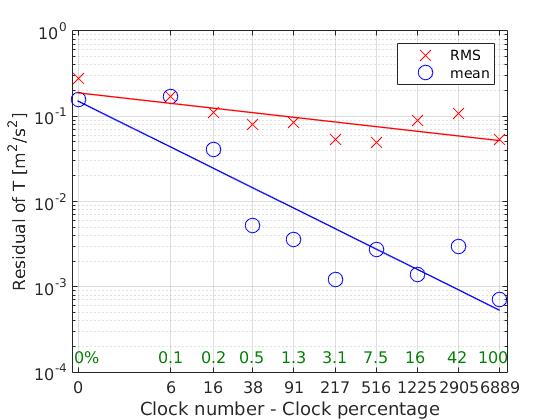}} \\
(b) Performance of the potential reconstruction (expressed by the rms and mean of 
    differences between the original potential on the regular grid and the reconstructed 
    one) wrt the number of clocks. 6989 gravity data have been used in this test. In green: number of clock data in terms of 
    percentage of~$\grdist$ data. \\
\end{tabular}
\caption{Contribution of clocks to the reconstruction of the disturbing potential in the Massif Central area.}
\label{fig:Contrib_Clocks_Auv}
\end{figure}

In more details, the main steps of this methodology are thus:
\begin{enumerate}
\item High spatial resolution 5-km step grids of gravity anomalies 
      $\Xref{\grdist}$ and the disturbing potential $\Xref{T}$ are generated with 
      the program GEOPOT~\citep{Smith_1998aa}, which allows to compute the gravity field related quantities at given locations by using mainly a geopotential model. These reference solutions, 
      see Figure~\ref{fig:Auv}, are obtained by using a state of the art geopotential 
      model \citep[EIGEN-6C4]{Forste_2014aa} up to degree and order 2000 (10~km resolution). 
      The long wavelengths of the gravity field covered by the satellites and longer than the extent of the local area are removed up to 200~km resolution, providing centered data or close to zero for the determination of local covariance function. The terrain effects are removed with the help of the topographic potential model dV\_ELL\_RET2012 
      \citep{Claessens_2013aa};

\item The distribution of gravimetric data is chosen to be the same than the control grid and contains 6989 location points. Several spatial distributions of clock, more and less dense, are generated by randomly sampling points from this regular grid, see Figure~\ref{fig:Contrib_Clocks_Auv} (top); and synthetic measurements $\Xsyn{\grdist}$ and $\Xsyn{T}$ are simulated on these points as previously;

\item A white noise is added to the simulated data $\Xsyn{\grdist}$ and~$\Xsyn{T}$, with a standard 
      deviation of 0.1~{m$^2$~$s^{-2}$} (i.e. 1~cm on the geoid) for clocks and 1~mGal for 
      gravimetric measurements;

\item The disturbing potential~$\Xest{T}$ is estimated from the 6989~synthetic measurements~$\grdist$ 
      only, and from the combination~$\Xsyn{T}$ and~$\Xsyn{\grdist}$ on the 5-km step grid using 
      the Least-Square Collocation (LSC) method \citep{Moritz_1980aa}. In this step, 
      we make an assumption on the gravity field regularity in the target area, 
      using a logarithmic 3D covariance function \citep{Forsberg_1987aa}. This 
      model has the advantage to provide the auto-covariances (ACF) and 
      cross-covariances (CCF) of the potential~$\Xsyn{T}$ and its derivatives in closed-form 
      expressions. Parameters of this model are adjusted on the empirical ACF 
      of~$\Xsyn{\grdist}$ with the program GPFIT \citep{Forsberg_2008aa}. Note that this 
      covariance function contains low frequencies that we have removed from 
      the data in step 1;

\item Finally, the potential recovery quality is evaluated for all the data 
      distribution sets and types of data by comparing the statistics of the 
      residuals between~$\Xest{T}$ and~$\Xref{T}$. 
\end{enumerate}

Figure~\ref{fig:Contrib_Clocks_Auv}(b) shows the residuals between the original 
and the reconstructed potential for denser and denser distributions of clocks. 
Clock number zero means that only gravity data have been used.

With no clock, the geoid is reconstructed with an accuracy of 3~cm in 
rms (or 0.3~{m$^2$~s$^{-2}$ in term of potential variations). When clocks with an accuracy of 1-cm for geoid height differences (or 0.1~m$^2$~s$^{-2}$) are added to the network, they improve 
the rms by several factors and reduce the bias up to 2--3~orders of 
magnitude. Also, we can see that it is not necessary to have a dense clock 
coverage, beyond 10 percent of the clock number of gravity points, to 
improve the determination of the disturbing potential. In this test, a coverage of around 
1--3~percent of the gravity data could be sufficient to reach the 
centimeter-level precision and improve greatly the bias. 

This simple example shows that clock-based geodetic observables provide useful 
information at spatial scales beyond those of the satellites for geopotential 
determination. The gain would be even larger in areas of higher relief. A more 
detailed study discussing the role of different parameters, such as the noise 
level in the data, effects of the resolution of gravity measurements and modeling errors can be found in \citep{Lion_2017aa}, in which they consider realistic spatial samplings. The addition of gravimetric data outside the modeling target area could also be considered in future works.

\section{Space-borne gravimetry with clocks}
\label{section:space-borne}

% {\it Contributions from Pieter and Dominic} \\

Sustained observation of the time varying gravity field of the Earth
with the highest possible precision and spatio-temporal resolution is
crucial for understanding mass transport in the Earth system, which
contributes to a better understanding of all kinds of geophysical
processes and their relation to climate change, geohazards, etc.
\citep{esa2015}. The US/German Gravity Recovery and Climate Experiment
(GRACE) has shown the tremendous potential of observing mass changes
due to all kinds of processes, including continental hydrology \citep{tapley2004},
ice melt in Greenland and West-Antarctica \citep{velicogna2006a,velicogna2006b},
and depletion of huge aquifer systems \citep{richey2015}. The importance of
sustained observation is reflected by the realization of the upcoming
GRACE Follow-On mission \citep{sheard2012}.

GRACE-type missions rely predominantly on so-called low-low
satellite-to-satellite tracking (ll-SST), where the distance between
two trailing satellites, typically with a distance of the order of 100-200
km, is measured with a precision of a few microns for microwave systems 
\citep{dunn2003}. This technique is especially suitable
for observing mass changes with a temporal resolution of the order of weeks
to months at spatial resolutions of at best a few hundreds of kms.
The European Space Agency (ESA) Gravity field and steady-state Ocean
Circulation Explorer (GOCE) used the concept of satellite gravity gradiometry
(SGG), where minute acceleration differences between adjacent capacitive
accelerometers (baseline $\approx$ 0.5 m) are observed \citep{drinkwater2007}.
This concepts allows the observation of the gravity field with a spatial
resolution of around 70-100 km, but is not well-suited for observing
temporal gravity field changes.

Sustaining and at the same time improving the observation of mass
transport through space-borne gravimetry has led to the selection
of GRACE Follow-on. This mission will not only carry a radiowave
instrument, but also a laser link for providing range observations
with a precision that is claimed to be a factor 20 better than the GRACE
microwave system \citep{jpl2017}. In addition, studies have been carried out that
demonstrate a great potential for enhancing both the temporal
and spatial resolution of flying multiple ll-SST missions
in parallel, and at the same time also enhancing the global
isotropy and homogeneity of gravity field solutions \citep{tas2010}.
However, there is a limit of what will be possible in terms
of spatial and temporal resolution when using ll-SST or
GOCE-type SGG. Therefore, new concepts are under study,
including the use of cold atom interferometry \citep{carraz2014},
and the use of ultra-precise space-borne atomic clocks
\citep{mayrhofer2012}. 

\subsection{Space-borne high-precision clocks}
\label{subsec:space-borne-clocks}

Here, we analyze the use of clock comparison between a LEO satellite and terrestrial ground stations for the purpose of determining the Earth's gravity field. The frequency of an electromagnetic signal transmitted from the ground and observed at the spacecraft will be shifted because of both the relative velocity and the difference in gravitational potential between the transmitter and receiver. By combining the measured frequencies, the following observable is obtained \citep{Kopeikin_2011_book}:
\begin{align}
\frac{f_{1}}{f_{2}}=\frac{\tau_{2}}{\tau_{1}}=\left(\frac{d\tau}{dT}\right)_{2}\frac{dT_{2}}{dT_{1}}\left(\frac{dT}{d\tau}\right)_{1}\label{eq:oneWayFrequencyTransfer}
\end{align}
where $T$ and $\tau$ denote the TCG coordinate time and observer proper time, respectively. The 2 and 1 subscripts denote the receiver (spacecraft) and transmitter (ground station). The terms $d\tau/dT$ are given to post-Newtonian order by Eq. (\ref{kop4}). The ratio of coordinate times (in TCG) can, to first order, be expressed as follows:
\begin{align} 
\frac{dT_{2}}{dT_{1}}=\frac{1-\frac{\hat{\mathbf{k}}\cdot\mathbf{v}_{2}}{c}}{1-\frac{\hat{\mathbf{k}}\cdot\mathbf{v}_{1}}{c}}\approx 1-\frac{\hat{\mathbf{k}}\cdot\mathbf{v}_{21}}{c}\label{eq:nonRelativisticDoppler}
\end{align}
where $\mathbf{v}_{2}$ and $\mathbf{v}_{1}$ denote the velocity vectors of receiver and transmitter (evaluated at $T_{2}$ and $T_{1}$), and $\mathbf{v}_{21}=\mathbf{v}_{2}-\mathbf{v}_{1}$. In the above, $\hat{\mathbf{k}}$ denotes the unit vector from transmitter position $\mathbf{r}_{1}$ to receiver position $\mathbf{r}_{2}$, so:
\begin{align}
\hat{\mathbf{k}}=\frac{\mathbf{r}_{2}(T_{2})-\mathbf{r}_{1}(T_{1})}{|\mathbf{r}_{2}(T_{2})-\mathbf{r}_{1}(T_{1})|}\label{eq:nonRelativisticDoppler2}
\end{align}
The relation in Eq. (\ref{eq:nonRelativisticDoppler}) is the so-called radial (or first-order) Doppler effect: the direct result of a change in transmission coordinate time on the reception coordinate time. In the following, we will assume that $d\tau/dT$ at the ground station can be modelled/measured to an uncertainty that is much better than for the space segment. The influence on the velocity of the spacecraft on its local $d\tau/dT$ is termed the quadratic (or second-order) Doppler effect.

For this application, we wish to retrieve the information on the Earth's gravitational potential $V_{E}$ from measurements of ${f_{2}}/{f_{1}}$. However, the measured frequency ratio encodes both velocity and potential differences. To map the influence of both spacecraft velocity uncertainties and the strength of the potential signature to uncertainties in the model for $\Delta(f_{2}/f_{1})$, we write for the first-order errors:

\begin{align}
\Delta\left(\frac{f_{2}}{f_{1}}\right)_{_{\text{radial Doppler}}}&=\left|\frac{d}{d\left(\hat{\mathbf{k}}\cdot\mathbf{v}_{21}\right)}\left(\frac{dT_{2}}{dT_{1}}\right)\right|\Delta\left(\hat{\mathbf{k}}\cdot\mathbf{v}_{21}\right)\approx\frac{\Delta\left(\hat{\mathbf{k}}\cdot\mathbf{v}_{21}\right)}{c}\label{eq:frequencyNonRelDopplerError}\\
\Delta\left(\frac{f_{2}}{f_{1}}\right)_{_{\text{quadratic Doppler}}}&=\left|\frac{d}{dv_{2}}\left(\frac{d{\tau}}{dT}\right)_{2}\right|\Delta\left(v_{2}\right)\approx\frac{v_{2}\Delta\left(v_{2}\right)}{c^{2}}\label{eq:frequencyRelDopplerError}\\
\Delta\left(\frac{f_{2}}{f_{1}}\right)_{_{\text{Einstein}}}&=\left|\frac{d}{dV_{E}}\left(\frac{d{\tau}}{dT}\right)_{2}\right|\Delta\left(V_{E}\right)\approx\frac{\Delta\left(V_{E}\right)}{c^{2}}\label{eq:frequencyPotentialError}
\end{align}
where the $\Delta\left(*\right)$ denotes the uncertainty in the parameter/measurement $*$. By combining Eq. (\ref{eq:frequencyPotentialError}), with Eqs. (\ref{eq:frequencyNonRelDopplerError}), and (\ref{eq:frequencyRelDopplerError}), respectively, we obtain the uncertainty in the potential determination $\Delta\left(V_{E}\right)$ induced by the velocity uncertainty as a result of the radial and quadratic Doppler effect, due to $\Delta\left(\hat{\mathbf{k}}\cdot\mathbf{v}_{21}\right)$ and $\Delta\left(v_{2}\right)$, respectively.

The influence of the radial Doppler effect is by far the greatest. The maximum allowable error in $\Delta\left(\hat{\mathbf{k}}\cdot\mathbf{v}_{21}\right)$ needed to obtain a given uncertainty $\Delta V_{E}$ for the potential is obtained from Eqs. (\ref{eq:frequencyNonRelDopplerError}) and (\ref{eq:frequencyPotentialError}):
\begin{align}
\Delta\left(\hat{\mathbf{k}}\cdot\mathbf{v}_{21}\right)=\frac{\Delta V_{E}}{c}
\end{align}
A potential height error $\Delta h_{g}$ corresponds to a potential error of $\Delta h_{g}\cdot g$, with $g$ the gravitational acceleration. Consequently, for $\Delta h_{g}$=1 cm, we get $\Delta\left(\hat{\mathbf{k}}\cdot\mathbf{v}_{21}\right)=\frac{0.01g}{c} \approx 3 \times 10^{-10}$ m/s or 0.3 nm/s (assuming a LEO satellite flying at 7.5 km/s with  $g \approx$ 9 m/s$^2$). This value is well below the orbit uncertainty of state-of-the-art orbit determination. However, it is possible to largely eliminate the radial Doppler shift by using a coherent two-way optical link \citep{djerroud2010,blanchet2001}. It is not possible, however, to eliminate the influence of the quadratic Doppler effect, since we cannot separate the influence of $v$ and $V_{E}$ in the equation for $(d\tau/dT)_{2}$ by combining a one- and two-way link. Again assuming an influence of maximum 1 cm on the geoid determination, the quadratic Doppler term results in a required spacecraft velocity uncertainty of better than $\frac{0.01 \times g}{7.5 \times 10^{3} } \approx 12 ~ \mu$m/s.

In light of the influence of the radial Doppler effect, it will be crucial to have a two-way link, in addition to the one-way frequency comparison. In the error simulations below, it is therefore assumed that such a two-way link is available.

\subsection{Accuracy assessment}
\label{subsec:sb_grav_results}

% Note: "nodal day", "ground track", and "repeat orbit" are well known 
% terms in Celestial Mechanics.
A hypothetical mission is defined where a satellite is flying in a low
Earth orbit carrying an ultra-precise clock that can be compared continuously
with perfect reference clocks. It is assumed that this satellite flies
in a circular orbit, where its projection on the surface of the Earth (the
so-called ground track) repeats after a certain number of nodal days. For such 
an observation geometry, use can be made of an efficient error propagation method that was introduced
in \citep{colombo1984}. This error propagation tool has been compared and
validated by comparison with robust numerical integration methods and
end-to-end simulations for several observation types, including orbit
perturbations, SGG and ll-SST \citep{schrama1991,visser2001}.

The satellite is assumed to fly in a polar orbit (inclination $i=90^{\circ}$) for
which the ground track repeats every month (31 days) in which the satellite
completes 497 orbital revolutions. The height above the Earth's surface
of the satellite is about 250 km. The space-borne clock is assumed
to have a frequency dependent instability equal to $30 \times 10^{-18}/\sqrt{\tau}$,
where $\tau$ represents the clock integration time. This instability is about
equivalent to a stability of $10^{18}$ for an integration time $\tau$ or 1000 s,
{\it cf.} \citep{giorgetta2013}. Assuming the clock instability to be the only 
source of measurement error $\Delta\left(f_{2}/f_{1}\right)$, Eq. (\ref{eq:frequencyPotentialError})
leads to the following error for the observation of the gravity potential:
\begin{eqnarray}
\frac{\Delta V_{E,\sigma(clock)} }{g} \approx 30 ~ \rm{cm}/\sqrt{\tau}
\label{eq:poterr_clock}
\end{eqnarray}
According to Eq.~(\ref{eq:frequencyRelDopplerError}), the velocity then has to be known with
a precision of about:
\begin{eqnarray}
\Delta v~=~360 ~ \mu \rm{m/s} /\sqrt{\tau}
\label{eq:poterr_vel}
\end{eqnarray}
Determining the LEO satellite velocity with such a precision is very challenging,
but might be feasible with future optical links \citep{chiodo2012}. The requirement
for the velocity determination is based on the assumption that the velocity
uncertainty causes an error in the potential that is equal or smaller than the measurement error due to clock instability.

The gravitational potential is expanded in terms of spherical harmonics. The uncertainty in the gravity field coefficients is then written as follows, where we neglect the uncertainty in $r$, $\lambda$ and $\phi$:
\begin{eqnarray}
\frac{ \Delta V_{E} }{g} ~=~ \frac{1}{g} \frac{\mu}{r} \{ 1 + \sum_{l=2}^{\infty} \sum_{m=0}^{l}
\left( \frac{a_e}{r} \right)^l ( \Delta \bar{C}_{lm} \cos m \lambda \nonumber \\
 + \Delta \bar{S}_{lm} \sin m \lambda ) {\bar{P}_{lm}} ( \sin \phi ) \}
\label{eq:potential_sh}
\end{eqnarray}
where $\mu$ is the gravity parameter (the product of the universal gravitational
constant $G$ and Earth's mass $M$), $a_e$ is the mean
% Note: "mean equatorial radius" is the correct terminology.
equatorial radius, $r, \phi, \lambda$ are the
spherical coordinates (radius, geocentric latitude, longitude)
denoting the location of the space-borne clock, $\bar{P}_{lm}$ is the
normalized Legendre polynomial of spherical harmonic degree $l$ and order $m$, and
$\bar{\Delta C}_{lm}, \bar{\Delta S}_{lm}$ represent the estimated gravity field coefficients. 

For a circular satellite orbit with repeating ground track, Eq.~(\ref{eq:potential_sh}) can be represented
by a Fourier series \citep{visser2003} and the method outlined in
\citep{colombo1984} can be used for predicting the gravity field retrieval
performance. The gravity field can be determined from time series of
observed or determined satellite velocities as well. In that case,
use is made of a linear orbit perturbation theory, where the associated 
equations can also be represented by a Fourier series \citep{schrama1991}.
Then also the method in \citep{colombo1984} can be used for a gravity
field retrieval accuracy assessment. Please note that
in case of deriving gravity from satellite velocity perturbations, it is
assumed that non-gravitational accelerations are known through e.g. observation
by precise accelerometers or absent due to a drag-free control system.

\begin{figure}[ht]
\begin{center}
\includegraphics[width=1.00\textwidth,angle=0]{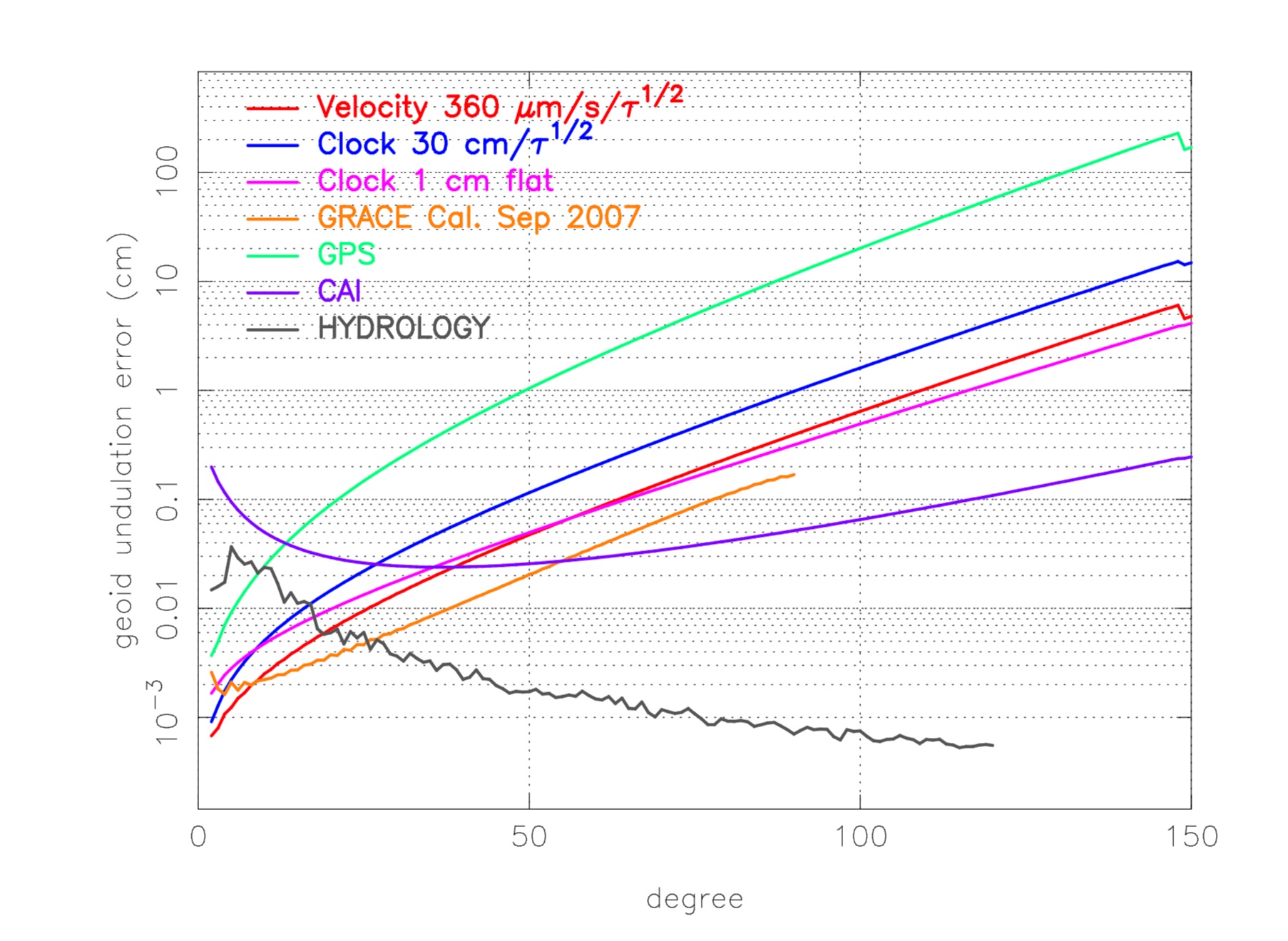}
\end{center}
\caption{Predicted gravity field retrieval errors for a LEO satellite in a 250-km
altitude circular orbit with monthly repeating ground track. The satellite
is equipped with a GPS receiver allowing 1-cm precision
orbit determination (GPS), a 3-dimensional gradiometer based on Cold Atom
Interferometry with flat noise of 0.4 mE (CAI), an ultra-precise clock
(Clock 30 cm/$\tau^{1/2}$), instrumentation for determining the 3-dimensional
velocity (Velocity 360 $\mu$m/s/$\tau^{1/2}$), and an ultra-precise clock resulting
in 1 cm potential height flat noise (Clock 1 cm flat). A typical curve for the
claimed GRACE performance (GRACE Cal Sep. 2007), and based
on a typical monthly gravity change due to continental hydrological
mass transport (HYDROLOGY) are included for reference.}
\label{fig:error_propagation}
\end{figure}

The predicted gravity field retrieval performances according to the error
spectra in Eq.~(\ref{eq:poterr_clock}) and (\ref{eq:poterr_vel})
are included in Figure~\ref{fig:error_propagation}. It can be observed
that the very precise knowledge of the satellite velocity leads to better
gravity field retrieval performance through classical orbit perturbation
analysis as compared to the post-Newtonian derivation of potential height
from clock observations.  For comparison, the predicted gravity field
performances for the same LEO satellite are included if equipped with
a high-quality Global Navigation Satellite System (GNSS) receiver allowing observation
of orbit position perturbations with 1 cm precision, and a gradiometer
based on cold atom interferometry (0.4 mE flat noise). We note that this assumption on the uncertainty is optimistic and does not reflect the state of the art for a space instrument, which is about one order of magnitude worse. With the used
assumptions, the ultra-stable clock does show a much better gravity
field performance than the GPS receiver, and also the gradiometer based
on cold atom inferometry up to spherical harmonic degree 30.
Figure~\ref{fig:error_propagation} also includes a performance prediction
for a clock that would have a flat error spectrum with a stability
of $10^{18}$ leading to 1 cm error in potential height. Such a - highly
speculative - performance would allow the observation of time variable
gravity with a precision that is relatively close to the currently
claimed performance level for GRACE Release 5 Level-2 products
(\citep{dahle2013}, see also Figure~\ref{fig:error_propagation}).

Being able to determine the 3-dimensional velocity of the LEO satellite
with a precision according to Eq.~(\ref{eq:poterr_vel}) leads to
a performance that is starting to become competitive with the currently
claimed GRACE performance and allows the observation of, for example,
gravity field changes due to continental hydrology (signal magnitude
also indicated in Figure~\ref{fig:error_propagation}, \citep{gruber2011}). In this preliminary analysis, we have not analyzed the possibilities to decouple the signature of the spacecraft's velocity and gravitational potential from the frequency comparison measurements. Such an approach, as is discussed by \cite{DirkxEtAl2016} for planetary laser ranging data, could be used to quantify more robustly the influence of velocity uncertainties on potential measurements using clocks. Conversely, such an analysis would quantify the degree to which the relativistic Doppler shift can be used to dynamically determine the spacecraft's state when making use of clock comparisons.

\section{Discussion and conclusions}
\label{section:conclusions}

Various geodetic applications of highly accurate and stable clocks at the $10^{-18}$ level for the relative frequency shift (corresponding to 1 cm in height) have been addressed. We discussed both the prospects of implementing in practice the new clock-based measurement concepts and their challenges. Our firm conclusion is that the quantum clocks and time metrology will provide us with unique opportunities to support and largely extend the tools for gravity field measurements and physical height determination based on the direct access to the gravity potential differences. 

We revisited the relevant time scales used in geodesy and showed the existence of growing inconsistencies in their definition and realization at a $10^{-18}$ accuracy level. This mainly concerns the use of the (non-constant) gravity potential value $W_0$ on the geoid, and its relation to the international time scales TT and TAI. In a foreseeable future, clarification and re-definition of the currently-used time concepts will be required to make them consistent with the highly-precise clock measurements.

We have provided a set of formulas to directly use the observed frequency shifts for the determination of physical height differences, where the relative clock measurements replace the differences of geopotential numbers. Applying this conceptually-new approach can facilitate or even fully resolve the problem of the systematic discrepancies existing between various height systems on the regional, national and intercontinental level. Clocks can connect tide gauges through satellite and/or fiber links providing the vertical datum of national height systems and help to determine the present offsets between them. Even discrepancies between independent geoid solutions stemmed from applying different classical techniques (GNSS/levelling versus gravimetric methods) can be determined and eliminated by making use of the precise chronometric measurements. These discrepancies are known to reaching centimeters to decimeters and can well be recovered already with the present-day clock technology. At the envisaged cm level of accuracy, time-variable parts affecting the clock measurements due to observer's height, position or Earth's density variations have to be considered. The most prominent effect is caused by the solid Earth tides but smaller variations, such as ocean tides or inaccurate knowledge of the Earth rotation parameters, play a significant role and are to be taken into consideration in the data processing algorithms.

The direct use of observed gravity potential values for a regional gravity field recovery has been studied as a test case in the Massif Central region crossing a number of mountain ranges. We have clearly demonstrated in our simulation scenario that adding the clock-based potential values to the existing data set would notably improve the final gravity field solution. The bias to a reference solution and total rms error could be reduced remarkably well, up to  a few orders of magnitude. Another important conclusion stemming from our simulations is  that in solving the problem of gravity field recovery it is not required to have a dense clock network, only a very few percent of clock measurements compared to the number of needed gravity data is sufficient.

The possible use of ultra-precise space-borne clocks has been revisited
for deriving potential height differences at satellite altitude and using these satellite data
for recovering the global gravity field. It has been shown that extremely
challenging requirements have to be met in order to use space-borne
clocks to successfully fulfill this task. In order to be able to exploit the gravity field
information content of clocks with a stability of $10^{-18}$, precise
reference clocks are needed along with two-way frequency transfer for eliminating
the Doppler effect correction errors. In addition, the position and velocity
of the clock need to be known with a very high precision. The current technological level of GNSS receivers
allow for determining the position with the required cm-level precision.
However, the velocity needs to be determined with such a high precision, that
the associated velocity perturbations caused by the gravity field anomalies might be used as well as a fundamental
observable for determining Earth's gravity field. Finally, the $10^{-18}$
clock stability needs to be achievable for very short integration times
of the order of seconds. Even if this would be possible and achieved,
the existing space geodesy concepts such as GRACE-type ll-SST lead to a better observability
of Earth's time variable gravity. However, our study does show that the frequency
transfer may be used to improve the determination of the spacecraft velocity. 

\begin{acknowledgement}
The International Space Science Institute (ISSI), is acknowledged for having 
provided the opportunity of presenting the work described in this paper at 
the Workshop on High Performance Clocks, 30 November - 4 December 2015 in 
Bern, Switzerland.
\end{acknowledgement}

\bibliographystyle{plainnat}      
% American Physical Society (APS) style, author-year citations
\bibliography{ISSI_Clock_Bibliography}

\end{document}